\documentclass[twocolumn]{article}
\usepackage{graphicx} 
\usepackage{amsmath,amssymb}

\usepackage[affil-it]{authblk}

\usepackage{abstract}

\usepackage{aas_macros}

\usepackage{hyperref}
\hypersetup{
    colorlinks=true,
    linkcolor=blue,
    filecolor=blue,  
    citecolor=blue,
    urlcolor=blue
    }

\usepackage[catalan,english]{babel}
\addto\captionscatalan{}

\title{\vspace{-4.2em} Sonification of gravitationally lensed gravitational waves / Sonificaci\'{o} de l'efecte de lent gravitat\`{o}ria en ones gravitacionals}

\author[1]{Helena Ubach \thanks{helenaubach@icc.ub.edu}}
\affil[1]{Institut de Ci\`{e}ncies del Cosmos (ICCUB), Universitat de Barcelona, Departament de F\'{i}sica Qu\`{a}ntica i Astrof\'{i}sica, Mart\'{i} i Franqu\`{e}s 1, E-08028 Barcelona, Spain}
\author[  ]{Jordi Espuny \thanks{jordi@zoom3.net}}
\affil[  ]{   }

\date{}

\begin{document}

\twocolumn[
    \maketitle
    \vspace{-4.6em}
    \begin{abstract}
      Gravitational waves are oscillations of space-time that are created, for example, in black hole mergers. If these waves travel through another massive astrophysical object, they will undergo an effect called gravitational lensing, that will distort and deflect them. This effect can create multiple images of the gravitational wave signal, and interference between them. In this document, we describe the sonification process of the gravitational lensing effect on gravitational waves. Sonification is the translation of data into sound. The wave nature of sound creates interference between waves in a natural way. This has allowed us to reproduce the interference produced by the superposition of gravitational wave images, characteristic of the gravitational lensing effect. 
      
      The results can be heard on the following websites:
            \begin{itemize}
          \item \href{https://zoom3.net/sonificacions/ona-gravitacional.html}{https://zoom3.net/sonificacions/ona-gravitacional.html} -- gravitational waves from the merger of two black holes
          \item \href{https://zoom3.net/sonificacions/ona-gravitacional-lent.html}{https://zoom3.net/sonificacions/ona-gravitacional-lent.html} -- gravitational waves affected by a gravitational lens, interactive website 
          \item \href{https://zoom3.net/sonificacions/ona-gravitacional-lent-exemples.html}{https://zoom3.net/sonificacions/ona-gravitacional-lent-exemples.html} -- gravitational waves affected by a gravitational lens, recorded examples
        \end{itemize}
    \end{abstract}
    \begin{otherlanguage}{catalan} 
    \begin{abstract}
      Les ones gravitacionals s\'{o}n oscil·lacions de l'espai-temps que es creen, per exemple, en col·lisions de forats negres. Si aquestes ones passen a trav\'{e}s d'un altre objecte astrof\'{i}sic massiu, patiran l'anomenat efecte de lent gravitat\`{o}ria, que les distorsionar\`{a} i deflectir\`{a}. Aquest efecte pot crear m\'{u}ltiples images del mateix senyal d'ones gravitacionals, i interfer\`{e}ncia entre elles. En aquest document, descrivim el proc\'{e}s de sonificaci\'{o} de l'efecte de lent gravitat\`{o}ria en ones gravitacionals. La sonificaci\'{o} \'{e}s la traducci\'{o} de dades a so. 
      La naturalesa d'ona del so crea interfer\`{e}ncia entre ones de manera natural. D'aquesta manera, hem pogut reproduir la interfer\`{e}ncia produ\"{i}da per la superposici\'{o} de les imatges de les ones gravitacionals, pr\`{o}pia de l'efecte de lent gravitat\`{o}ria.

      Els resultats de la sonificaci\'{o} es poden escoltar a les seg\"{u}ents p\`{a}gines web:
      \begin{itemize}
          \item \href{https://zoom3.net/sonificacions/ona-gravitacional.html}{https://zoom3.net/sonificacions/ona-gravitacional.html} -- ones gravitacionals provinents de la fusi\'{o} de dos forats negres
          \item \href{https://zoom3.net/sonificacions/ona-gravitacional-lent.html}{https://zoom3.net/sonificacions/ona-gravitacional-lent.html} -- ones gravitacionals amb efecte de lent gravitat\`{o}ria, web interactiva 
          \item \href{https://zoom3.net/sonificacions/ona-gravitacional-lent-exemples.html}{https://zoom3.net/sonificacions/ona-gravitacional-lent-exemples.html} -- ones gravitacionals amb efecte de lent gravitat\`{o}ria, exemples gravats
        \end{itemize}
    \end{abstract}
    \end{otherlanguage}
    \vspace{\baselineskip}
]
\saythanks

\clearpage

\section{Introduction}

In scientific research, the most usual way to present information is in a visual way, for instance, through plots and figures. However, other types of representations are also possible and they can provide a complementary point of view. One of such representations is sonification of scientific data, which allows us to represent the data through sound.

\begin{figure*}
    \centering
    \includegraphics[width=2\columnwidth]{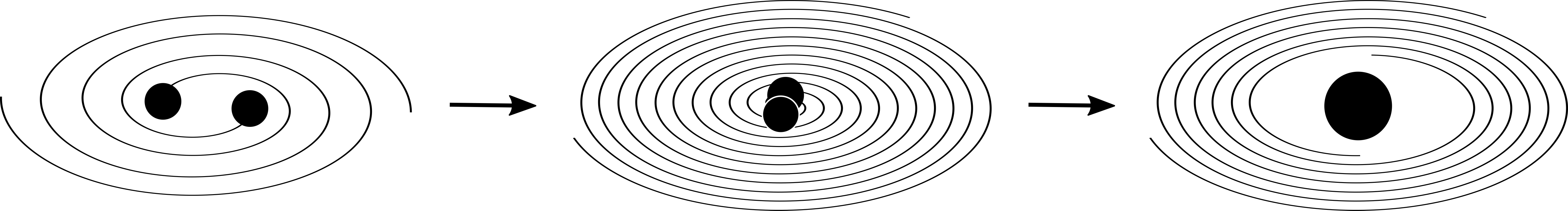}
    \caption{Black hole merger sequence. First, black holes orbit around each other, losing energy in gravitational waves and getting close. As they get closer, the frequency increases (Eq.~\eqref{eq:f-tau-eng}). Finally, they end up merging, forming a more massive black hole. When the final black hole relaxes, gravitational waves are not emitted anymore.
    }
    \label{fig:BH-merger-eng}
\end{figure*}

In this article, we present the sonification of gravitationally lensed gravitational waves. Gravitational waves are oscillations of space-time, created by the acceleration of very massive astrophysical objects, such as binary black hole systems which end up merging (Fig.~\ref{fig:BH-merger-eng}). These waves have been observed recently by detectors LIGO, Virgo and KAGRA \cite{LIGO18-GWTC1,LIGO20-GWTC2,LIGO21-GWTC3,venumadhav-20,zackay-21}. If gravitational waves encounter a massive astrophysical object while travelling through the cosmos, the gravity of this object distorts the waves, creating an effect known as gravitational lensing. Therefore, if these lensed waves arrive at the detectors, the signal will be different than a signal which has not been distorted.

Some sonifications of gravitational waves already exist, both with real events \cite{LVK-sonif-1,LVK-sonif-2,LVK-sonif-3,LVK-sonif-4} and all kinds of simulated gravitational waves: with eccentricity, spin, or binary system precession effects \cite{sounds-spacetime}. There are also some sonifications for possible sources of both LIGO and a future detector, LISA \cite{sounds-lisa}. There is even a game, ``Black Hole Hunter", based on the sonification of gravitational wave signals and how it helps in finding them hidden inside the noise \cite{bh-hunter}. 

To our knowledge, gravitationally lensed gravitational waves have not been sonified yet. In this work, we present their sonification for the first time. Although our purpose is mainly illustrative (for educational purposes), we have pursued physical rigor throughout the work.

\subsection{What is sonification?}

Sonification is a set of techniques that consist in transforming inaudible events or phenomena into audible sounds. This process is done with a specific and well defined purpose. For example, the 12 clock bell strokes indicate a given moment in the day through a sound. In this example, both phenomena are of a very different nature: time has been translated into sound. Here, sound and the relation it establishes with a moment in the day is totally arbitrary.

Sometimes, the phenomenon we want to sonify is already a wave, but an inaudible one. For example, seismic waves registered by a seismograph are mechanical waves like sound, but they cannot be heard. In this case, the sonification process requires a change in scale that adjusts the frequency and amplitude of the waves, so that they can enter the audible range.

With gravitational waves, the process is similar to the previous one. In this case, it is not a mechanical wave, so the sonification changes its nature, but it maintains some of its initial properties. As a consequence, the resulting sound has a sufficient fidelity that allows its comprehension and analysis.

\subsection{Sonification in science}

In the last decades, sonification has been used to interpret some physical phenomena from a complementary point of view. Even though sonification is used in outreach or with educational purposes, there are also some cases in which it has had a notable importance in science. If we focus in physics, we can find some examples. The Geiger counters generate sounds to warn when radiation crosses them (radioactivity) \cite{sonification-report}. Another case is the cosmic microwave background, which was detected as a persistent sound in a radio antenna. In the same sense, signals coming from pulsars also can be heard as radio pulses \cite{pulsar1,pulsar2}. 

Compared to the visualisation of data, the perception of sound could also enable the detection of hidden signals inside noise, such as when we listen to a conversation in the middle of a noisy crowd. A recurrent problem in astronomy is to detect hidden signals inside noise, such as gravitational waves \cite{bh-hunter,zanella-22}. In the reference \cite{zanella-22}, they present a more extensive repository of sonifications in science, and some others in other topics \cite{repositori-sonif}.

\section{Physics fundamentals}

\subsection{Gravitational waves}

Gravitational waves are oscillations of space-time itself. They are created when very massive astrophysical objects are accelerated \cite{einstein-gw-16}. For example, they could be emitted in a binary system of two black holes, that is, two black holes that are orbiting each other, as shown in Fig.~\ref{fig:BH-merger-eng}. We are focusing here in black holes, but they could also be other types of compact objects, such as neutron stars. Stellar-mass black holes have masses of $5$ to $100$ solar masses, $M = (5-100) \, M_\odot$, while neutron stars are lighter, $M \sim (1-3) \, M_\odot$.

\begin{figure}
    \centering
    \includegraphics[width=\columnwidth]{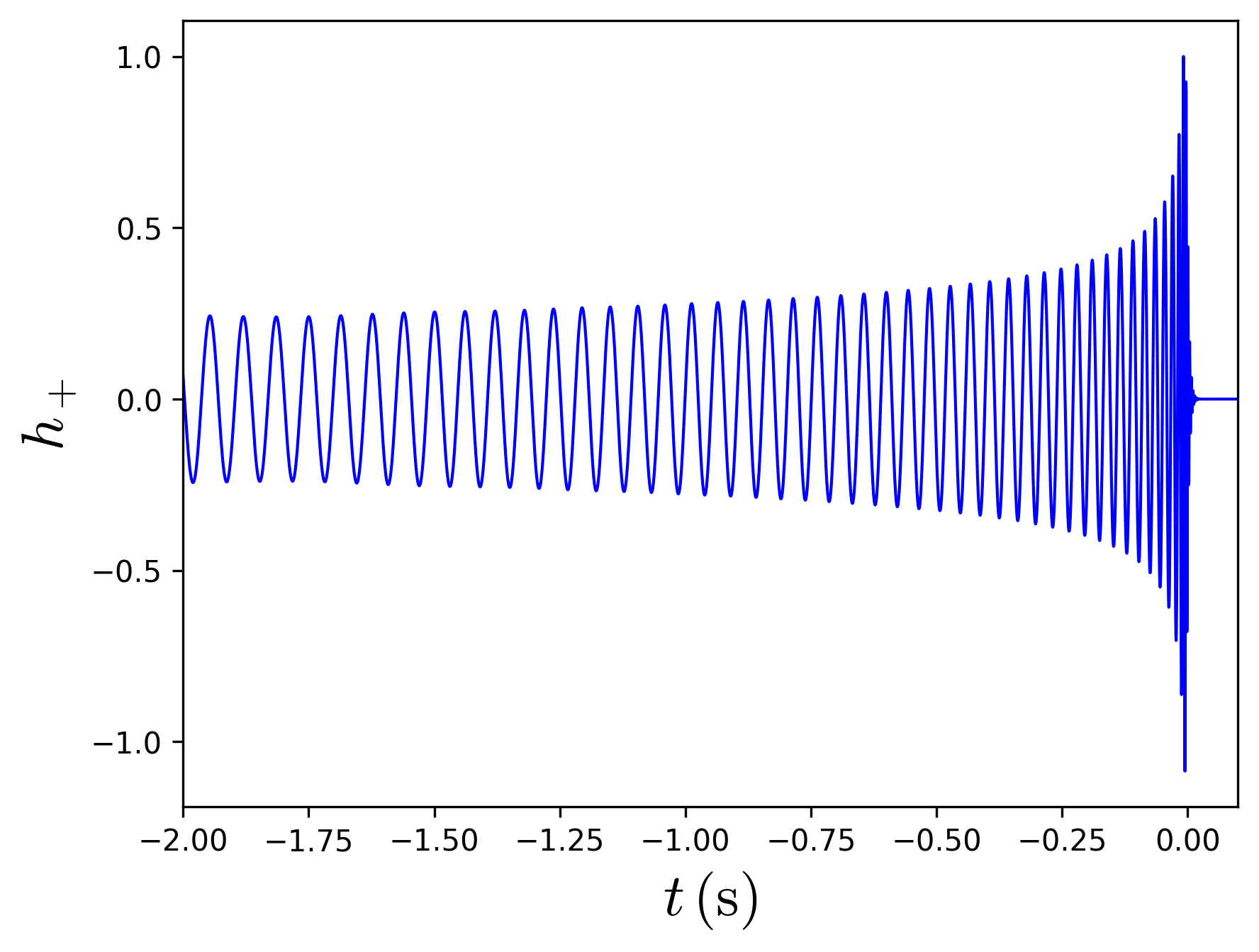}
    \caption{Space-time deformation due to gravitational waves from a black hole merger, represented through $h_+$ as a function of time $t$. The first part, before the amplitude maximum, corresponds to the approach between the black holes. It can be expressed with Eq.~\eqref{eq:h-t-eng}. The amplitude maximum corresponds to the merger: here, it is necessary to use calculations with General Relativity. Finally, the remnant black hole relaxes (``ringdown" phase), emitting the last gravitational waves that are rapidly attenuated. In this case, we represent the merger of two black holes of $30$ solar masses, but this figure represents qualitatively the shape of the signal for the other masses as well. Here $h_+$ is not to scale, the usual values would be $h_+\sim 10^{-21}$.}
    \label{fig:h-t-unlensed-eng}
\end{figure}

Gravitational wave signals coming from such binary systems have been detected with LIGO, Virgo and KAGRA detectors \cite{LIGO18-GWTC1,LIGO20-GWTC2,LIGO21-GWTC3,venumadhav-20,zackay-21}. When the two black holes orbit each other and emit gravitational waves, they lose energy due to that emission, causing them to come closer together (they fall onto each other due to gravity) and end up merging. The merger product is another, more massive, black hole. Once they have merged and the system relaxes, it does not emit gravitational waves anymore. The intensity (and amplitude) of gravitational waves is maximum at the time of merger. This type of signal (merger of two black holes) has the shape shown in Fig.~\ref{fig:h-t-unlensed-eng}. Once emitted, these gravitational waves travel through space-time in all directions at the speed of light. They will also pass through the Earth, where they can be detected.

The characteristics of gravitational waves depend mainly on the black hole masses\footnote{They also depend on other characteristics like spin, the binary system's eccentricity or the precession. Their effects, however, are only relevant when these characteristics are very significant (for example, a high eccentricity). For simplicity, we do not take into account the effects from these characteristics here.}. The characteristic mass is called \textit{chirp mass}, $\mathcal{M}_c$. It is a combination of the two black hole masses, $m_1$ and $m_2$ \cite{maggiore-07}: 

\begin{equation}
\mathcal{M}_c \equiv \frac{(m_1 m_2)^{3/5}}{(m_1+m_2)^{1/5}}.
\label{eq:mchirp-eng}
\end{equation}
If the masses are equal, the chirp mass will be very close to the mass of one of the objects, $\mathcal{M}_c = 2^{-1/5} \, m_1 \approx 0.87\, m_1$.

The chirp mass $\mathcal{M}_c$ determines the frequency evolution of gravitational waves. When the black holes are approaching and orbiting each other, the frequency of the gravitational waves they are creating increases with time \cite{maggiore-07}:
\begin{equation}
f(\tau) = \frac{1}{\pi}
\left(
\frac{5}{256} \frac{1}{\tau}
\right)^{3/8} 
\left(
\frac{G \mathcal{M}_c}{c^3} 
\right)^{-5/8},
\label{eq:f-tau-eng}
\end{equation}
where $\mathcal{M}_c$ is the chirp mass (defined in Eq.~\eqref{eq:mchirp-eng}), $G$ is the gravitational constant and $c$ is the speed of light in vacuum (also constant). The variable is $\tau = t_{\rm coal}-t$, where $t_{\rm coal}$ is the time of merger and $t$ is the time variable, taken from an origin. Therefore, $\tau$ represents the time left before the merger. At the time of merger, $\tau\rightarrow 0$ and $f\rightarrow \infty$, but it is impossible that the frequency could reach infinite values. What happens in reality is that, shortly before the merger, the black holes touch, and they are so close that Eq.~\eqref{eq:f-tau-eng} is not valid any more: gravity is very strong and it is necessary to use the complete equations of General Relativity. In these equations, frequency reaches a value, approximately 
\cite{berti-09}
\begin{equation}
f_0 \simeq 1.207 \cdot 10^4 \, {\rm Hz} 
\left( \frac{M_\odot}{M_S} \right),
\label{eq:freq-ms-eng}
\end{equation}
where $M_S$ corresponds to the mass of the black hole resulting from the merger, $M_S \approx m_1+m_2$ \footnote{In fact, $M_S$ does not coincide exactly with the sum of the black hole masses ($m_1+m_2$), that is the reason why we have written it as an approximation. The approximation is good enough to do a qualitative study. The fact that the masses do not coincide is precisely because energy is lost through gravitational waves: part of the black holes' mass $m_1$, $m_2$ is converted into pure energy.}. 

The smaller the mass, the higher the frequency. Conversely, the larger the mass, the lower the frequency. Even though the physical process is different, it can be compared to a musical instrument, where smaller instruments tend to have a higher pitch (higher frequency) while the larger instruments tend to have a lower pitch (lower frequency).

The amplitude of gravitational waves is measured through the deformation of space-time that is detected at the Earth, $h$. The amplitude also increases with time. If we take one of the polarisations, $+$\footnote{Taking one polarisation or another does not matter qualitatively. It only matters when the inclination (orientation) of the orbital plane is taken into account, but here we take only one specific value. For practical purposes, it only changes the overall gravitational wave amplitude.}, the amplitude $h_+$ can be written as \cite{maggiore-07}:
\begin{equation}
h_+(t) = \frac{1}{d}
\left(
\frac{G \mathcal{M}_c}{c^2} 
\right)^{5/4} 
\left(
\frac{5}{c \tau} 
\right)^{1/4} 
\left(
\frac{1+\cos^2{\iota}}{2} 
\right)
\cos{[\Phi(\tau)]},
\label{eq:h-t-eng}
\end{equation}
where $d$ is the distance to the source. $\iota$ is the inclination of the source's orbital plane, that we take for example as $\cos^2{\iota}=1$\footnote{The inclination is constant during the signal's duration and does not change the result qualitatively. The only effect is in increasing or decreasing a bit the global amplitude.}. Here we use $\tau = t_{\rm coal}-t$ again. In the same way as the frequency, even if $h_+$ reaches infinity for $\tau\rightarrow 0$, in reality the amplitude will never be infinite.

The phase $\Phi(\tau)$ that appears in Eq.~\eqref{eq:h-t-eng} is defined as 
\begin{equation}
\Phi(\tau) = -2 
\left(
\frac{5 G \mathcal{M}_c}{c^3} 
\right)^{-5/8}
\tau^{5/8} + \Phi_0.
\label{eq:phi-tau-eng}
\end{equation}
$\Phi_0$ is the initial phase, and it can be taken as an arbitrary value without changing the behaviour, therefore we take $\Phi_0=0$. The dependence of the frequency with time (Eq.~\eqref{eq:f-tau-eng}) is already included inside the phase (Eq.~\eqref{eq:phi-tau-eng}).

The sonification of gravitational waves emitted by the merger of two astrophysical objects (specifically, two black holes) can be heard in \href{https://zoom3.net/sonificacions/ona-gravitacional.html}{https://zoom3.net/sonificacions/ona-gravitacional.html}. The website is interactive and allows you to choose the chirp mass you would like to listen.

\subsection{Gravitational lensing effect}
\label{fonament-lent-eng}

When these gravitational waves are travelling, after being emitted, they could encounter another massive astrophysical object (such as another black hole or a galaxy) before arriving at the Earth. Since the gravity of this object curves space-time, according to the theory of General Relativity \cite{einstein-15}, gravitational waves will get distorted and their path will be deflected. This distortion and deflection effect is called the gravitational lensing effect. If any of the gravitational wave signals undergoes this effect and we detect it, its shape will be different than that of Fig.~\ref{fig:h-t-unlensed-eng}. The gravitational lensing effect is rare (it would affect approximately one every $1000$ signals in current observations \cite{ng-18,li-18,oguri-18,mukherjee-21,wierda-21}) and has not been observed in the detectors yet \cite{hannuksela-19, LVK-O3a-lensing, LVK-O3ab-Lensing}. Despite not having detected a lensed event yet, we can model what would be observed when we detect one.

The distortion will be different depending on the following parameters:

\begin{itemize}

\item the mass of the lens $M_L$, 

\item the frequency of the gravitational wave $f$, that will depend on the masses of the source black holes ($m_1$ and $m_2$), as we described in Eqs.~\eqref{eq:f-tau-eng},\eqref{eq:freq-ms-eng}.

\item the misalignment of the source respect to the lens and the observer, represented by the parameter $y$. When $y=0$, the source, the lens and the observer are perfectly aligned. As $y$ increases, they progressively misalign and the distortion due to the lensing effect decreases. $y$ is defined as 
\begin{equation}
y=\frac{d_L}{d_S}\frac{\eta}{R_{\rm E}}.
\end{equation}
It is a geometrical factor that contains a combination of distances and characteristic scales. As it can be seen in 
Fig.~\ref{fig:lens-geom-eng}, $d_L$ is the distance to the lens, $d_S$ is the distance to the source, $\eta$ is the transversal misalignment distance of the source and $R_{\rm E}$ is a characteristic scale of the system (known as Einstein radius)\footnote{$\displaystyle{R_{\rm E}=\sqrt{\frac{4 G M_L}{c^2} \frac{d_L d_{LS}}{d_S}}}$, known as the Einstein radius, is the characteristic scale of the gravitational lens. $d_{LS}$ is the distance between the lens and the source (as shown in Fig.~\ref{fig:lens-geom}).}. To practical effects, it is useful to imagine that $d_L$, $d_S$ and $R_{\rm E}$ are fixed, and $\eta$ gets modified. $y$ is directly proportional to $\eta$.

\begin{figure}
    \centering
    \includegraphics[width=\columnwidth]{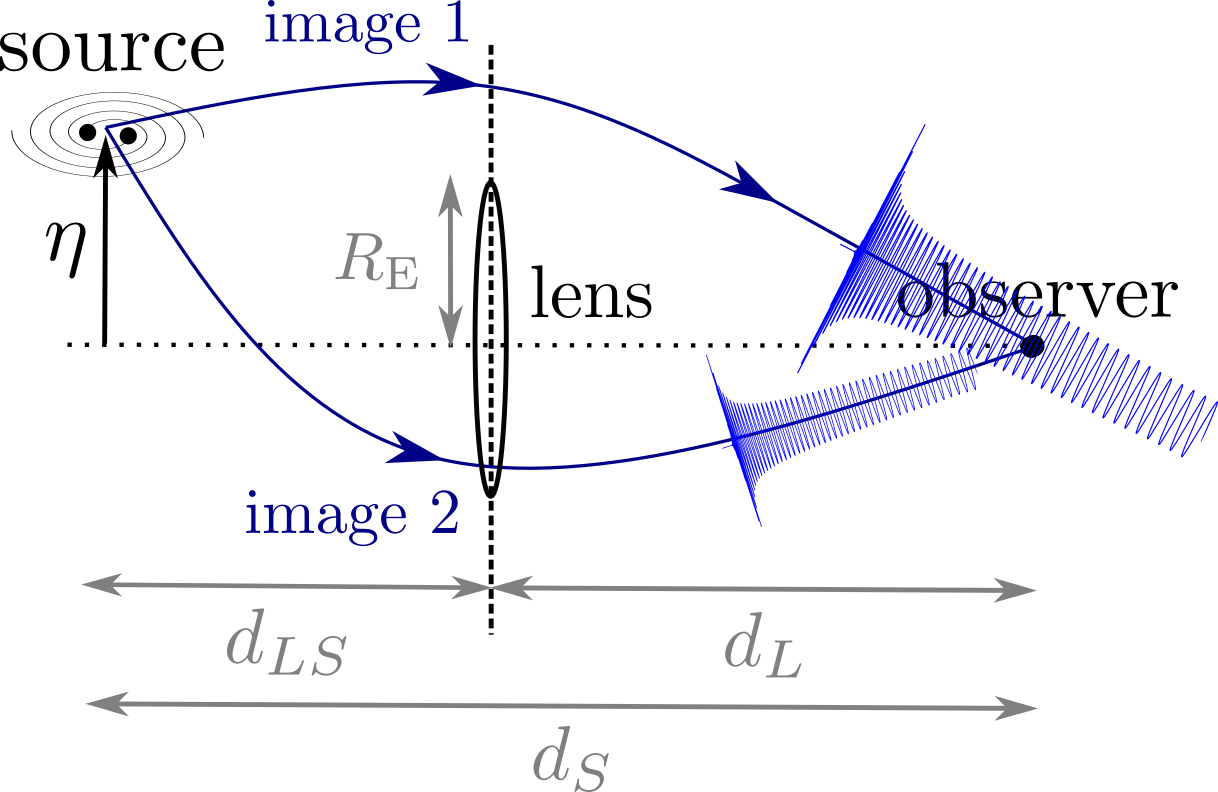}
    \caption{Diagram of the gravitational lensing effect. Gravitational waves from the source travel through a gravitational lens (another massive astrophysical object), they get distorted and deflected, and arrive at the observer (the detector on the Earth). In this case, the blue curved lines represent the trajectory of two images that could be formed (in the ``point mass" model), in the geometric optics approximation.}
    \label{fig:lens-geom-eng}
\end{figure}

It is interesting that in many cases the product $f\cdot M_L$ acts as an independent variable\footnote{In some studies, a ``dimensionless frequency" is defined, proportional to $f\cdot M_L$.}. Therefore, for example, decreasing $M_L$ would cause the same effect as decreasing $f$ would.

\end{itemize}

When the signal is distorted, it can appear magnified and/or as multiplied images, similarly at what occurs in optical systems such as magnifying glasses, lenses and mirrors. There are mainly three types of effects: (1) amplification of the signal, (2) interference between multiple separate images of the same source, when they overlap, (3) multiple separate images of the same source, when they do not overlap.

We consider the gravitational lens as a ``point mass", that is, concentrated in a point in space. For this lens model, two images are formed. For other more complex lens models, more than two images could appear.

The type of effects that we would observe depends on the parameters of the lens and the geometry of the system. We take $y$ and the product \mbox{$f\cdot M_L$} as parameters. Let us see what would happen if we consider one of the parameters as a constant and we modify the other one:

\begin{itemize}

\item First, let us take the parameter $y$ as a constant. If the product $f\cdot M_L$ is small, there will only be amplification  (Fig.~\ref{fig:h-t-lensed-eng} (a)). If this product increases (for example, increasing the mass of the lens, or increasing the frequency of the source), then there are two overlapped images, interfering between them (Fig.~\ref{fig:h-t-lensed-eng} (b)). If $f\cdot M_L$ increases even more, these images will be seen as two separate signals (Fig.~\ref{fig:h-t-lensed-eng} (c)).

\begin{figure}
    \centering
    \includegraphics[width=\columnwidth]{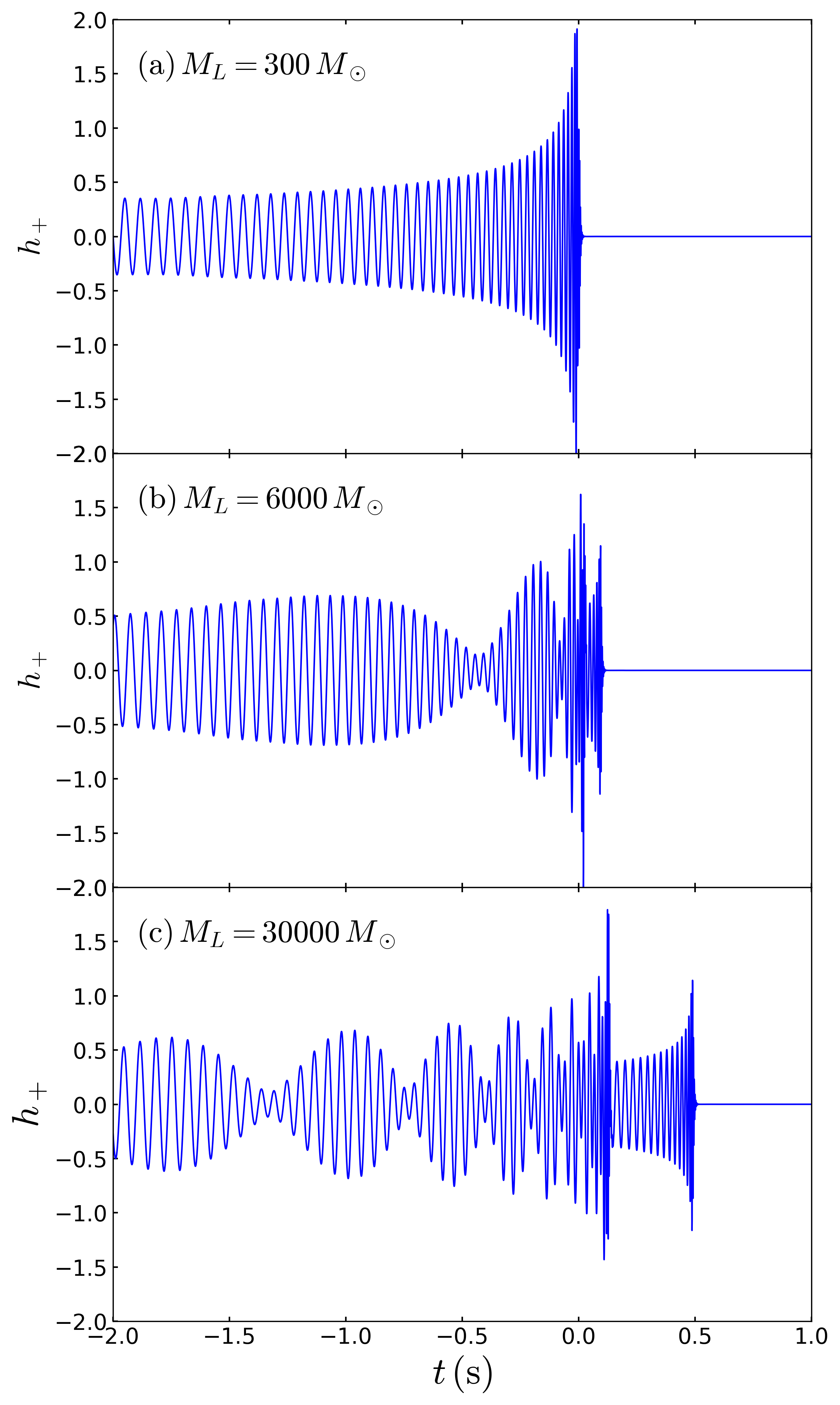}
    \caption{Space-time deformation due to gravitationally lensed gravitational waves from a black hole merger, represented through $h_+^L$ as a function of time $t$. The three subfigures show different cases, with the same value for $y$ in all of them ($y=0.3$). The mass of the lens is changed: (a) $M_L = 300 \, M_\odot$, (b) $M_L = 6000 \, M_\odot$, (c) $M_L = 30000 \, M_\odot$. We have used some of the cases which we can play with in the website \href{https://zoom3.net/sonificacions/ona-gravitacional-lent.html}{https://zoom3.net/sonificacions/ona-gravitacional-lent.html}.}
    \label{fig:h-t-lensed-eng}
\end{figure}

\item On the other hand, we now take the mass of the lens and the frequency of the source as constants (therefore, $f\cdot M_L$ will also be a constant). If the parameter $y$ is very small, the images are not well defined and there will only be amplification. If $y$ increases a bit, the images will start to be defined, but they will be close and overlapped, and there will be interference between them. For larger values of $y$, the images will be further apart.

Additionally, the amplification of the images, as we will see  ($\sqrt{\mu_1}$ and $\sqrt{\mu_2}$, Eqs.~\eqref{eq:mu1-eng},~\eqref{eq:mu2-eng}), only depends on the parameter $y$. The smaller $y$, the ``stronger" will be the final signal (combination of the two images) in general. 

\end{itemize}

At the website \href{https://zoom3.net/sonificacions/ona-gravitacional-lent.html}{https://zoom3.net/sonificacions/ona-gravitacional-lent.html} you can play with the parameters $\mathcal{M}_c$, $M_L$ and $y$ to check the different cases that have been described.

\subsubsection{Geometric optics approximation: multiple images}

The distortion can be quantified by the so-called ``transmission factor" $F$ \cite{born-99,takahashi-03}. Gravitational waves with the lensing effect can be described as $h_+^{L} = F \cdot h_+$ (in frequency space). When the images are defined, and they appear either separated or interfering, we can use an approximation, ``geometric optics" (GO) \cite{schneider-92,nakamura-deguchi-99,takahashi-03}:
\begin{align}
h_+^{L} 
&= F_{GO} \cdot h_+ \nonumber \\
&= h_+ \sqrt{\mu_1}\,e^{i \phi_1} + h_+ \sqrt{\mu_2} \,e^{i \phi_2-i \pi/2}.
\label{eq:F_GO_PML-eng}
\end{align}
As it can be seen, the signal with the lens effect, $h_+^L$, is the sum of two signals: the first signal is like the original $h_+$, but the amplitude is modified with the factor $\sqrt{\mu_1}$ (Eq.~\eqref{eq:mu1-eng}); the second signal with the factor $\sqrt{\mu_2}$ (Eq.~\eqref{eq:mu2-eng}). Furthermore, both come with different phases (one with $\phi_1$ and the other one with $\phi_2-\pi/2$). If we represent the amplitude in terms of the relative phase between the images,
\begin{equation}
h_+^{L} 
= \left(h_+ \sqrt{\mu_1} + h_+ \sqrt{\mu_2} \,e^{i 2 \pi \Delta t - i \pi/2} \right) e^{i \phi_1},
\label{eq:GO-relative-eng}
\end{equation}
where we have written $\phi_2-\phi_1 = 2\pi \Delta t$. This means that the second signal comes with a phase difference of \mbox{$2 \pi \Delta t - \pi/2$} respect to the first one, which would translate into a time delay $\Delta t$ \footnote{Additionally, there is the term $-\pi/2$ which comes from the topology of the function that determines the time delay.}. Therefore, the second signal arrives some time $\Delta t$ later than the first one. This temporal separation between images is given by
\begin{equation}
\Delta t \simeq 4\cdot 10^{-5} \,y \, \left(
\frac{M_L}{M_\odot}\right) \, {\rm s},
\label{eq:time-delay-eng}
\end{equation}
in units of seconds (s).
As the mathematical expression shows, the time separation (or delay) between the images increases when (i) $y$ increases, (ii) $M_L$ increases. 

The amplification/magnification of each image only depends on the parameter $y$. The magnification of the first and the second image are, respectively, \cite{schneider-92}:
\begin{equation}
\sqrt{\mu_{1}}=\displaystyle{\sqrt{\frac{1}{4} 
\left( \frac{y}{\sqrt{y^2+4}} + \frac{\sqrt{y^2+4}}{y} + 2 \right) }},
\label{eq:mu1-eng}
\end{equation}
\begin{equation}
\sqrt{\mu_{2}}=\displaystyle{\sqrt{\frac{1}{4} 
\left( \frac{y}{\sqrt{y^2+4}} + \frac{\sqrt{y^2+4}}{y} - 2 \right) }}.
\label{eq:mu2-eng}
\end{equation}
However, the geometric optics approximation is not valid in the case where there is only amplification. In this case, there is diffraction and the images are not totally defined. Figure \ref{fig:h-t-lensed-eng} takes into account the effect of diffraction. For simplicity, however, for the sonification we will work as if the images were always defined, using Eq.~\eqref{eq:F_GO_PML-eng}. Therefore, we need to take that into account when interpreting the results. The geometric optics approximation overestimates the amplification in the case where there is only amplification (Fig.~\ref{fig:h-t-lensed-eng} (a)), where diffraction is important. On the other hand, it does not affect the other cases, interference and separate images, where the geometric optics approximation is valid. If we would like to take into account the diffraction effect correctly, we should use a more elaborate mathematical formalism. For more details, the reader is referred to the reference \cite{BU-2022}.

\subsubsection{Wave effects: interference and diffraction}

Apart from producing multiple images, the gravitational lens can produce effects associated with the wave nature of gravitational waves. These ``wave effects" are mainly interference and diffraction.

Interference (between waves) is produced when two or more waves encounter each other, which creates a distortion in the resulting wave. For example, it can be seen when two stones are dropped to a pond: the interference will be produced when the waves encounter each other. Similarly, we have seen that the gravitational lens can create multiple images of the same source. Even if the images are separated in time by $\Delta t$, since they have a certain duration, they could overlap. When these images overlap, there is interference between them. The interference effect is led by the relative phase between the waves: from Eq.~\eqref{eq:GO-relative-eng}, we can see that the relative phase difference between the images in our study would be $2\pi \Delta t - \pi/2$. In some regions, interference will be constructive (the amplitudes of the images are summed) or destructive (the amplitudes of the images are subtracted). Consequently, this will produce an interference pattern as beats (as we can see in Fig.~\ref{fig:h-t-lensed-eng} (b),(c)).

Diffraction, on the other hand, is also an interference phenomenon, but that of multiple partial waves (here we talk about waves in general, not images). When $\Delta t$ is very small\footnote{Formally, we can take the condition of ``very small $\Delta t$" as $\Delta t \lesssim 1/f$.}, the two images are still not well defined. In this case, the contribution comes from partial waves (``parts of the waves") from the source, distorted by the lens. Diffraction can create an amplification of the signal (Fig.~\ref{fig:h-t-lensed-eng} (a)). However, if we compare it with the amplification produced by the effect of overlapping images, the amplification from diffraction is a bit more suppressed. To generate Fig.~\ref{fig:h-t-lensed-eng} (a) we have taken into account the effect of diffraction. On the other hand, to do the sonification, we have used the model of two well defined images (geometric optics). This implies that for small values of the product $f\cdot M_L\cdot y$ (when there is only amplification), the amplification we see will be overestimated respect to the case where diffraction is taken into account. 

\subsection{Comparison between gravitational waves and sound}

Even though we are sonifying gravitational waves, that does not mean that we can hear them in reality. The sound waves are mechanical (pressure) waves that can only travel through a material medium (air/gas, but also through liquids and solids). Their velocity also depends on the medium where they propagate through. Gravitational waves, on the other hand, are waves of space-time itself, and propagate both through vacuum and any medium at the speed of light. In that case, it is not a wave that propagates through space-time, but the deformation of space-time itself that propagates: when a gravitational wave gets through us, we also get distorted because we are inside space-time (we should not worry though, because the effect is so tiny that we do not feel it). Despite the difference between the two types of waves (sound and gravitational waves), they both possess a wave behaviour that allows us to do a precise enough comparison.

\section{Sonification process}

\subsection{Gravitational waves}

\subsubsection{First test: empirical correspondence}

The sonification process has been made in different stages. In a first test, the sonification has been made in an approximate way. To do that, we have relied on the shape of the plots made from the equations. The purpose of this part of the process is to obtain a first sound result as fast as possible.

The gravitational wave shown in Fig.~\ref{fig:h-t-unlensed} has a exponential shape. Its frequency also increases exponentially. When we generate a generic exponential wave, since it does not come from data from the equations, we can freely choose all the values, until we get a similar wave: in duration of the wave, the initial and final frequencies, as well as the phase shift due to the gravitational lensing effect (next section, Sec.~\ref{sonif-lent-eng}). Through several iterations, we have been able to replicate a similar pattern.

\subsubsection{Realistic procedure}
\label{procediment-realista-eng}

In a second step, we used realistic equations that describe gravitational waves correctly. We generated the frequency variation behaviour using Eq.~ \eqref{eq:f-tau-eng}.

Finally we used Eq.~\eqref{eq:h-t-eng}, that generates the complete gravitational wave: frequency $f$ and amplitude $h_+$ as a function of time $t$, where we used $t = t_{\rm coal}-\tau$. To avoid infinity at $\tau\rightarrow 0$, we cut the equation at $\tau = 0.0001 \, {\rm s}$. With that, we are taking the first part of the gravitational wave as a signal (from the time where the black holes are orbiting until when they touch and would then merge).  
Eq.~\eqref{eq:h-t-eng}, apart from taking into account the amplitude variation $h_+(t)$, also takes into account the frequency variation (Eq.~\eqref{eq:f-tau-eng}) through the phase $\Phi(\tau)$. The frequency variation can be observed through the separation between the oscillations in Fig.~\ref{fig:h-t-unlensed-eng}: they start further apart (lower frequency) and progressively get closer (higher frequency).

The amplitude of all the signals has been scaled to make them audible at a comfortable level. When you listen to them, please be careful not to have the volume too high: even though at the beginning the signal could sound weaker, its intensity will increase exponentially.

The frequencies of the gravitational waves we consider are very close to audible sound frequencies.
However, in some cases the obtained frequencies remain outside the audible range, which is theoretically from 20 Hz to 20000 Hz\footnote{Another limitation is the frequency response of the computer speakers, which can be between 80 Hz and 15000 Hz.}. Gravitational waves from more massive black holes will have lower frequencies. The earlier frequencies will have so low frequency that we will not be able to hear them, we will only hear the final part of the signal (where the frequencies have increased sufficiently to enter the audible range). To be able to hear the whole wave, we need to modify the obtained frequencies. Thus, we have increased the global frequency of all the signals by $300$ Hz.

\subsection{Gravitational lensing effect}
\label{sonif-lent-eng}

To simulate the gravitational lensing effect, we have directly used the realistic effect, using the geometric optics approximation. The gravitational lensing effect can create two images of the same source
\footnote{As we have seen in Sec.~\ref{fonament-lent-eng}, there is a region where diffraction is important, where the images are still not well defined. For the sonification process, however, we use the geometric optics approximation (Eq.~\eqref{eq:F_GO_PML}) for simplicity.}. Depending on the lens parameters, $M_L$ and $y$, it could be that the images arrive with a shorter or longer time difference $\Delta t$ (Eq.~\eqref{eq:time-delay}). 
For the sonification, instead of representing the amplitude  $h_+^L$ with Eq.~\eqref{eq:F_GO_PML-eng} directly (which could be done), we have decided to overlap two copies of the same signal to simulate the lensing effect. Each signal has been scaled with the corresponding factor ($\sqrt{\mu_1}$ (Eq.~\eqref{eq:mu1-eng}) for the first one, $\sqrt{\mu_2}$ (Eq.~\eqref{eq:mu2-eng}) for the second one), and the second image has been delayed by $\Delta t$. The extra phase $-\pi/2$ of the second image has also been taken into account.

\subsubsection{Interference pattern}

To create the interference, we can proceed in two ways:

1 - Reproducing the interference of waves in an acoustic way, by using the left and right speakers to reproduce the same wave, with a phase difference, corresponding to two images separated by a delay $\Delta t$. The interference is produced in a natural way as a real acoustic phenomenon, instead of being computed. The result, therefore, depends on the separation between the left and right speaker, and also on the position where they are heard from.

2 - If we want to obtain a more accurate result, which also does not depend on the configuration of the reproduction and listening system, the resulting wave can be created digitally.

It is important to note that the interference pattern is generated in a natural way, by overlapping the two time-shifted waves. This occurs thanks to the wave nature of sound signals, which also have a phase. The relative phase between the two signals will reproduce the same pattern that the lensing effect would create. 

\subsubsection{Frequency shift}

As we mentioned in Sec.~\ref{procediment-realista-eng}, we need to modify the obtained frequencies to make the signal completely audible. There are some strategies to do that.

The first strategy consists in modifying the reproduction velocity of the obtained sound. In this case, every time that the reproduction velocity is multiplied by a factor of two, the frequency increases by a factor of two, but the signal duration is also reduced by two.

If we want to keep the duration of the original signal and the interference pattern, there is another way, based in the Fourier transform (FFT, Fast Fourier Transform). This allows us to shift the frequency components towards the higher frequency part of the spectrum. Finally, we can use the inverse Fourier transform (IFFT) to recover the wave. This operation does not modify the amplitude of the wave and the interference pattern is kept unchanged. This operation can digitally distort the original wave, introducing undesirable artifacts, so the frequency shift cannot be too large.

\section{Results}

\subsection{Sonification of gravitational waves from the merger of two black holes}

The sonification of gravitational waves from the merger of two black holes can be found in the website:

\href{https://zoom3.net/sonificacions/ona-gravitacional.html}{https://zoom3.net/sonificacions/ona-gravitacional.html}

You can choose the chirp mass $\mathcal{M}_c$ of the source you would like to hear, by writing it in the text box. Then, press the left button ``Send" to create the shape of the signal with the mass you have chosen, which will be displayed in the image. Finally, press the right button ``Play sound" to hear the corresponding sonification.

Please be careful with the loudness and the pitch of the sound. The default starting chirp mass $\mathcal{M}_c=30\, M_\odot$ is a middle safe frequency and one can gradually increase or decrease from there.

\subsection{Sonification of the gravitational lensing effect on gravitational waves}

The sonification of the gravitational lensing effect on gravitational waves from the merger of two black holes can be found in the websites:

\href{https://zoom3.net/sonificacions/ona-gravitacional-lent.html}{https://zoom3.net/sonificacions/ona-gravitacional-lent.html}

\href{https://zoom3.net/sonificacions/ona-gravitacional-lent-exemples.html}{https://zoom3.net/sonificacions/ona-gravitacional-lent-exemples.html}

In the first website, you can choose the values of the chirp mass $\mathcal{M}_c$ of the source, the lens mass $M_L$ and the parameter $y$. The shape of the resulting gravitational wave will be displayed below. The interactive sonification, however, is still under development.

In the second website, you can hear some of the recorded sonifications while the first website is under development.

\section{Discussion}

The wave nature of sound allows us to reproduce the gravitational lensing effect on gravitational waves. Not only the individual images, but also the interference between images, which is produced in a natural way. We can check that, even if the sound and gravitational waves have a different nature, they share behaviours which we can use for the sonification.

For the sonification process, we have taken the freedom of slightly shifting the range of frequencies to make the signals audible (towards higher frequencies). To avoid modifying other characteristics of the signal, specially the interference pattern, we have used a method with the Fourier transform (FFT) and its inverse (IFFT).

We have used the geometric optics approximation, which considers that the images are always well defined. This allows us to create two images separately and overlap them, creating the interference. As discussed, this approximation loses its validity when the time shift is very small ($\Delta t \lesssim 1/f$), where diffraction appears. The main effect of using the approximation, and not taking into account diffraction, is the overestimation of the amplitude in the case where the lens effect only produces the amplification of the signal. For the other cases (multiple images and interference), the approximation is precise enough.

\subsection{Future steps}

The websites are in continuous development. We will continue developing and updating them to improve their presentation and to explain their operation. 

Now that we have done the sonification of the gravitational lensing effect on gravitational waves, the next step would be to explore new sonifications from a more artistic perspective. The final step of this exploration and creation process is expected to be a public exhibition in Barcelona, with an educational focus.

\section*{Acknowledgements} 

We are grateful to the organizers of the initiative ``Sounds and Astrophysics" (``Arts $\&$ Science"), which has encouraged the collaboration between the Faculty of Fine Arts at the University of Barcelona and the Institute of Cosmos Sciences (ICCUB). Arnau Rios and Jordi Miralda provided suggestions for the initial idea of the project. We particularly thank Anna Argudo for her implication and lead in the organisation of the project.

To generate Figs.~\ref{fig:h-t-unlensed-eng} and \ref{fig:h-t-lensed-eng} we have used the PyCBC software (\href{https://zenodo.org/records/10473621}{https://zenodo.org/records/10473621}). To create Figs.~\ref{fig:BH-merger-eng} and \ref{fig:lens-geom-eng}, we have used Inkscape. To create the sonifications, we have used SuperCollider.

Helena Ubach is supported by the FI-SDUR predoctoral grant from the Generalitat de Catalunya. 

\setcounter{section}{0} 
\setcounter{figure}{0}
\setcounter{equation}{0}

\clearpage

\begin{otherlanguage}{catalan}

\section{Introducci\'{o}}

En recerca cient\'{i}fica, la manera m\'{e}s habitual de presentar la informaci\'{o} \'{e}s visualment, per exemple a trav\'{e}s de gr\`{a}fics. Tanmateix, altres tipus de representacions tamb\'{e} s\'{o}n possibles i poden aportar un punt de vista complementari. Una d'aquestes representacions \'{e}s la sonificaci\'{o} de dades cient\'{i}fiques, que permet representar les dades a trav\'{e}s del so.

En aquest article, presentem la sonificaci\'{o} d’ones gravitacionals deflectides per l’efecte de lent gravitat\`{o}ria. Les ones gravitacionals s\'{o}n oscil·lacions de l’espai-temps, creades per l’acceleraci\'{o} d’objectes astrof\'{i}sics molt massius, com ara sistemes binaris de forats negres que acaben col·lisionant (Figura \ref{fig:BH-merger}). Aquestes ones s’han pogut mesurar recentment amb els detectors LIGO, Virgo i KAGRA \cite{LIGO18-GWTC1,LIGO20-GWTC2,LIGO21-GWTC3,venumadhav-20,zackay-21}. Si aquestes ones es troben amb un objecte astrof\'{i}sic massiu en viatjar a trav\'{e}s del cosmos, la gravetat d’aquest objecte les distorsiona, creant l’efecte de lent gravitat\`{o}ria. D’aquesta manera, si aquestes ones deflectides arriben als detectors, el seu senyal ser\`{a} diferent que el d’una ona no distorsionada.

Ja existeixen algunes sonificacions d’ones gravitacionals, tant dels esdeveniments reals \cite{LVK-sonif-1,LVK-sonif-2,LVK-sonif-3,LVK-sonif-4} com de simulacions d’ones gravitacionals amb diferents efectes: d'excentricitat, d’esp\'{i} o  de precessi\'{o} del sistema binari \cite{sounds-spacetime}. Tamb\'{e} hi ha sonificacions de fonts tant de LIGO com d'un futur detector, LISA \cite{sounds-lisa}. Fins i tot s'ha creat un joc, ``Black Hole Hunter", basat en la sonificaci\'{o} de senyals d'ones gravitacionals i com la sonificaci\'{o} ens permet trobar-los amagats dins el soroll \cite{bh-hunter}. 

Les ones gravitacionals que pateixen l’efecte de lent gravitat\`{o}ria encara no han estat sonificades, segons tenim coneixement. En aquest treball presentem la seva sonificaci\'{o} per primer cop. Encara que el nostre prop\`{o}sit \'{e}s principalment il·lustratiu (per a finalitats did\`{a}ctiques), hem intentat mantenir el rigor f\'{i}sic al llarg de l'estudi.

\begin{figure*}
    \centering
    \includegraphics[width=2\columnwidth]{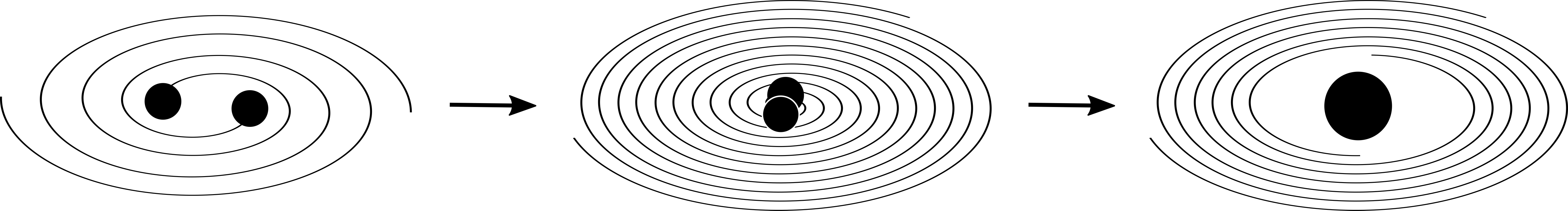}
    \caption{Seq\"{u}\`{e}ncia de fusi\'{o} de dos 
    forats negres. Primer orbiten un al voltant de l'altre, perden energia en ones gravitacionals i s'acosten. A mesura que s'acosten, la freq\"{u}\`{e}ncia va augmentant (Eq.~\eqref{eq:f-tau}). Finalment acaben fusionant-se, formant 
    un forat negre m\'{e}s massiu.
    Quan el forat negre resultant es relaxa,
    ja no s'emeten m\'{e}s ones gravitacionals.}
    \label{fig:BH-merger}
\end{figure*}

\subsection{Qu\`{e} \'{e}s la sonificaci\'{o}?}

La sonificaci\'{o} s\'{o}n un conjunt de t\`{e}cniques que consisteixen en transformar esdeveniments o fen\`{o}mens inaudibles en sons audibles. Aquest proc\'{e}s es fa amb un prop\`{o}sit espec\'{i}fic i ben definit. Per exemple, les 12 campanades del rellotge de l'esgl\'{e}sia ens indiquen un moment concret del dia mitjançant el so. En aquest exemple, els dos fen\`{o}mens s\'{o}n de naturalesa molt diferent: s'ha tradu\"{i}t el temps a so. Aqu\'{i}, el so i la relaci\'{o} que s'estableix amb un moment del dia \'{e}s del tot arbitr\`{a}ria. 

A vegades, el fenomen que es vol sonificar ja \'{e}s una ona, per\`{o} no audible. Per exemple, les ones s\'{i}smiques registrades per un sism\`{o}graf s\'{o}n ones mec\`{a}niques com el so, per\`{o} no es poden escoltar. El proc\'{e}s de sonificaci\'{o}, en aquest cas, implica un canvi d'escala que ajusta la freq\"{u}\`{e}ncia i l'amplitud de les ones per a qu\`{e} entrin en el rang audible. 

Amb les ones gravitacionals el proc\'{e}s \'{e}s similar a l'anterior. En aquest cas no es tracta d'una ona mec\`{a}nica, per tant la sonificaci\'{o} en canvia la seva natura, per\`{o} mant\'{e} algunes de les propietats inicials. D'aquesta manera, el so resultant t\'{e} un grau de fidelitat suficient que permet la seva comprensi\'{o} i estudi.

\subsection{Sonificaci\'{o} en ci\`{e}ncia}

Al llarg de les \'{u}ltimes d\`{e}cades, la sonificaci\'{o} s'ha fet servir per a interpretar alguns fen\`{o}mens f\'{i}sics des d'un punt de vista complementari. 
Tot i que habitualment s'utilitza en divulgaci\'{o} o amb finalitats educatives, tamb\'{e} hi ha hagut alguns casos en que ha tingut una import\`{a}ncia notable en ci\`{e}ncia. Si ens fixem en la f\'{i}sica, podem trobar alguns exemples. Els comptadors Geiger generen sons per avisar quan hi travessa radiaci\'{o} (radioactivitat) \cite{sonification-report}. D'altra banda, el fons c\`{o}smic de microones es va detectar com un so persistent en una antena de r\`{a}dio. En el mateix sentit, els senyals procedents dels p\'{u}lsars tamb\'{e} es poden escoltar com a pulsacions en r\`{a}dio \cite{pulsar1,pulsar2}. 

En comparaci\'{o} amb la visualitzaci\'{o} de les dades, la percepci\'{o} del so tamb\'{e} permetria detectar senyals amagades en soroll, com quan escoltem una conversa concreta enmig d'una multitud de gent parlant. Sovint en astronomia apareix el problema de detectar senyals amagades en soroll, com les ones gravitacionals  \cite{bh-hunter,zanella-22}. A la refer\`{e}ncia \cite{zanella-22} es presenta una recopilaci\'{o} m\'{e}s extensiva de sonificacions en ci\`{e}ncia, aix\'{i} com algunes en altres disciplines \cite{repositori-sonif}.

\section{Fonaments f\'{i}sics}

\subsection{Ones gravitacionals}

Les ones gravitacionals s\'{o}n oscil·lacions del propi espai-temps. Es creen quan objectes astrof\'{i}sics molt massius s\'{o}n accelerats \cite{einstein-gw-16}. Per exemple, es poden emetre en un sistema binari de dos forats negres, \'{e}s a dir, dos forats negres que orbiten un al voltant de l’altre, com es mostra a la Fig.~\ref{fig:BH-merger}. Encara que aqu\'{i} ens centrem en forats negres, tamb\'{e} poden ser altres objectes compactes, com estrelles de neutrons. Els forats negres estel·lars tenen masses de $5$ a $100$ masses solars, $M = (5-100) \, M_\odot$, mentre que les estrelles de neutrons s\'{o}n m\'{e}s lleugeres, $M \sim (1-3) \, M_\odot$. 

Els senyals d’ones gravitacionals provinents d’aquest tipus de sistemes binaris s’ha pogut detectar amb els detectors LIGO, Virgo i KAGRA \cite{LIGO18-GWTC1,LIGO20-GWTC2,LIGO21-GWTC3,venumadhav-20,zackay-21}. Quan els dos forats negres orbiten i emeten ones gravitacionals, perden energia degut a aquesta emissi\'{o}, fent que cada cop s’apropin m\'{e}s (caiguin l’un sobre l’altre per gravetat) i acabin fusionant-se. El resultat de la fusi\'{o} \'{e}s un altre forat negre, m\'{e}s massiu. Un cop s’han fusionat i el sistema es relaxa, ja no emet m\'{e}s ones gravitacionals. La intensitat (i amplitud) de les ones gravitacionals \'{e}s m\`{a}xima durant el moment de la fusi\'{o}. Per ara, els senyals detectats pertanyen a la part final de l’\`{o}rbita i durant la fusi\'{o}. Aquest tipus de senyal (fusi\'{o} de dos forats negres) t\'{e} la forma que es pot veure en la Fig.~\ref{fig:h-t-unlensed}. Un cop emeses, aquestes ones gravitacionals viatgen a trav\'{e}s de l’espai-temps en totes direccions a la velocitat de la llum. Tamb\'{e} travessaran la Terra, on es poden detectar. 

\begin{figure}
    \centering
    \includegraphics[width=\columnwidth]{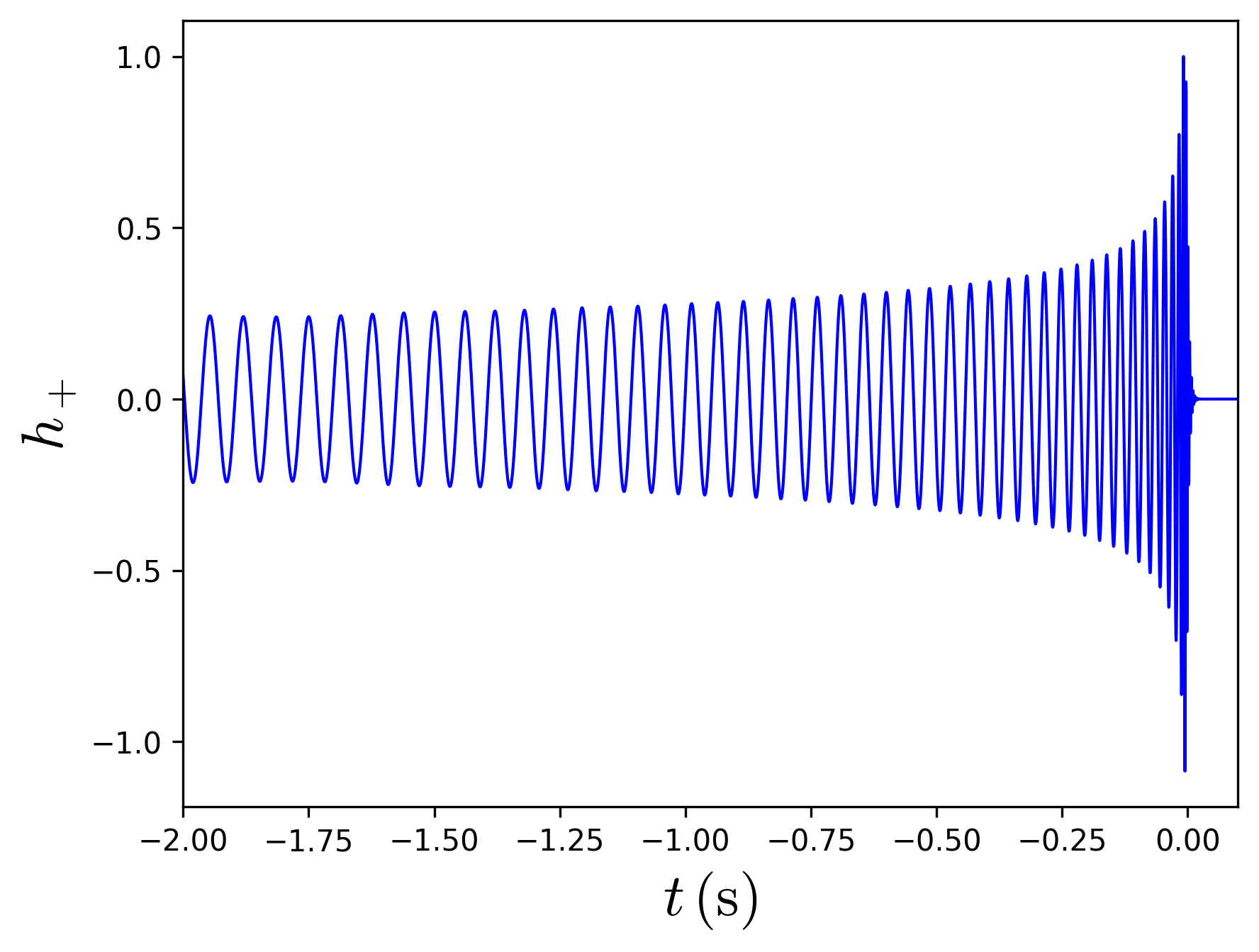}
    \caption{Deformaci\'{o} de l'espai-temps deguda a les ones gravitacionals provinents de la fusi\'{o} de dos forats negres, representat a trav\'{e}s de $h_+$ en funci\'{o} del temps $t$. La primera part, abans del m\`{a}xim d'amplitud, correspon a l'acostament entre forats negres. Es pot expressar amb l'Eq.~\eqref{eq:h-t}. El m\`{a}xim d'amplitud correspon a la fusi\'{o}: aqu\'{i} \'{e}s necessari utilitzar c\`{a}lculs amb Relativitat General. Finalment, el forat negre final es relaxa (fase ``ringdown"), produint unes \'{u}ltimes ones gravitacionals que s'atenuen r\`{a}pidament. En aquest cas, representem la fusi\'{o} de dos forats negres de $30$ masses solars, per\`{o} aquesta figura tamb\'{e} representa qualitativament la forma del senyal per a altres masses. Aqu\'{i} $h_+$ no est\`{a} a escala, els valors habituals serien $h_+\sim 10^{-21}$.}
    \label{fig:h-t-unlensed}
\end{figure}

Les caracter\'{i}stiques de les ones gravitacionals depenen principalment de les masses dels forats negres\footnote{Tamb\'{e} depenen d'altres caracter\'{i}stiques com l'esp\'{i}, l'excentricitat del sistema binari o la precessi\'{o}. Els seus efectes, per\`{o}, nom\'{e}s s\'{o}n rellevants quan aquestes caracter\'{i}stiques s\'{o}n molt significatives (per exemple, una alta excentricitat). Per simplicitat, aqu\'{i} no tenim en compte els efectes d'aquestes altres caracter\'{i}stiques.}. La massa caracter\'{i}stica s'anomena \textit{massa de chirp}, $\mathcal{M}_c$. \'{E}s una combinaci\'{o} de les masses dels dos forats negres, $m_1$ i $m_2$ \cite{maggiore-07}: 
\begin{equation}
\mathcal{M}_c \equiv \frac{(m_1 m_2)^{3/5}}{(m_1+m_2)^{1/5}}.
\label{eq:mchirp}
\end{equation} 
Si les masses s\'{o}n iguals, la massa de chirp ser\`{a} semblant a la massa d'un dels objectes, $\mathcal{M}_c = 2^{-1/5} \, m_1 \approx 0.87\, m_1$.

La massa de chirp $\mathcal{M}_c$ determina l'evoluci\'{o} de la freq\"{u}\`{e}ncia de l'ona gravitacional.
Quan els forats negres s'estan acostant i orbitant, la freq\"{u}\`{e}ncia de les ones gravitacionals que creen augmenta amb el temps \cite{maggiore-07}:
\begin{equation}
f(\tau) = \frac{1}{\pi}
\left(
\frac{5}{256} \frac{1}{\tau}
\right)^{3/8} 
\left(
\frac{G \mathcal{M}_c}{c^3} 
\right)^{-5/8},
\label{eq:f-tau}
\end{equation}
on $\mathcal{M}_c$ \'{e}s la massa de chirp (definida a l'Eq.~\eqref{eq:mchirp}), $G$ \'{e}s la constant de la gravitaci\'{o} universal i $c$ \'{e}s la velocitat de la llum en el buit (tamb\'{e} constant). La variable \'{e}s $\tau = t_{\rm coal}-t$, on $t_{\rm coal}$ el moment de la fusi\'{o} i $t$ la variable del temps, pres des d'un punt d'origen. Per tant, $\tau$ representa el temps que queda per la fusi\'{o}. En el moment de la fusi\'{o}, $\tau\rightarrow 0$ i $f\rightarrow \infty$, per\`{o} \'{e}s impossible que la freq\"{u}\`{e}ncia pugui arribar a ser infinita. El que ocorre en realitat \'{e}s que, poc abans de la fusi\'{o}, els forats negres es toquen, i estan tan propers que l'Eq.~\eqref{eq:f-tau} ja no \'{e}s v\`{a}lida: la gravetat \'{e}s molt forta i cal fer servir les equacions completes de la Relativitat General. En aquestes equacions, la freq\"{u}\`{e}ncia assoleix un valor, aproximadament 
\cite{berti-09}
\begin{equation}
f_0 \simeq 1.207 \cdot 10^4 \, {\rm Hz} 
\left( \frac{M_\odot}{M_S} \right),
\label{eq:freq-ms}
\end{equation}
on $M_S$ correspon a la massa del forat negre resultant de la fusi\'{o}, $M_S \approx m_1+m_2$ \footnote{En realitat, $M_S$ no coincideix exactament amb la suma de la massa dels forats negres ($m_1+m_2$), per aix\`{o} hem escrit que \'{e}s una aproximaci\'{o}. L'aproximaci\'{o} ja ens va b\'{e} per fer un an\`{a}lisi qualitatiu. El fet que no coincideixin les masses \'{e}s precisament perqu\`{e} es perd energia en forma d'ones gravitacionals: una part de la massa dels forats negres $m_1$, $m_2$ es converteix en pura energia.}. 

Com m\'{e}s petita \'{e}s la massa, m\'{e}s alta \'{e}s la freq\"{u}\`{e}ncia. Com m\'{e}s gran \'{e}s la massa, m\'{e}s baixa \'{e}s la freq\"{u}\`{e}ncia.
Tot i que el proc\'{e}s f\'{i}sic \'{e}s diferent, es pot fer el s\'{i}mil amb un instrument musical, on els instruments m\'{e}s petits acostumen a ser m\'{e}s aguts (freq\"{u}\`{e}ncia m\'{e}s alta), mentre que els instruments m\'{e}s grans acostumen a ser m\'{e}s greus (freq\"{u}\`{e}ncia m\'{e}s baixa). 

L'amplitud de l'ona gravitacional es mesura mitjançant la deformaci\'{o} de l'espai-temps que es detecta a la Terra, $h$. L'amplitud tamb\'{e} augmenta amb el temps. Si prenem una de les polaritzacions, $+$\footnote{Prendre una polaritzaci\'{o} o altra no varia qualitativament. Nom\'{e}s t\'{e} import\`{a}ncia quan tenim en compte la inclinaci\'{o} (orientaci\'{o}) del pla de l'\`{o}rbita dels forats negres, per\`{o} aqu\'{i} nom\'{e}s agafem un valor qualsevol. A efectes pr\`{a}ctics nom\'{e}s varia la amplitud de l'ona gravitacional global.}, l'amplitud $h_+$ es pot escriure com  \cite{maggiore-07}:
\begin{equation}
h_+(t) = \frac{1}{d}
\left(
\frac{G \mathcal{M}_c}{c^2} 
\right)^{5/4} 
\left(
\frac{5}{c \tau} 
\right)^{1/4} 
\left(
\frac{1+\cos^2{\iota}}{2} 
\right)
\cos{[\Phi(\tau)]},
\label{eq:h-t}
\end{equation}
on $d$ \'{e}s la dist\`{a}ncia a la que es troba la font. $\iota$ \'{e}s la inclinaci\'{o} del pla de l'\`{o}rbita de la font, que prendrem per exemple com a $\cos^2{\iota}=1$\footnote{La inclinaci\'{o} \'{e}s constant durant la durada del senyal i no varia qualitativament el resultat. L'\'{u}nic efecte que t\'{e} \'{e}s augmentar o disminuir una mica l'amplitud global.}. Aqu\'{i} tornem a utilitzar $\tau = t_{\rm coal}-t$. De la mateixa manera que amb la freq\"{u}\`{e}ncia, encara que $h_+$ es faci infinita per $\tau\rightarrow 0$, en realitat l'amplitud no es fa mai infinita.

La fase $\Phi(\tau)$ que apareix a l'Eq.~\eqref{eq:h-t} est\`{a} definida com 
\begin{equation}
\Phi(\tau) = -2 
\left(
\frac{5 G \mathcal{M}_c}{c^3} 
\right)^{-5/8}
\tau^{5/8} + \Phi_0.
\label{eq:phi-tau}
\end{equation}
$\Phi_0$ \'{e}s la fase inicial i es pot agafar de manera arbitr\`{a}ria sense variar el comportament, per tant prenem $\Phi_0=0$. La depend\`{e}ncia de la freq\"{u}\`{e}ncia amb el temps (Eq.~\eqref{eq:f-tau}) ja est\`{a} considerada dins la fase (Eq.~\eqref{eq:phi-tau}).

La sonificaci\'{o} d'ones gravitacionals emeses per fusions de dos objectes astrof\'{i}sics (en concret, dos forats negres) es pot escoltar a la p\`{a}gina web \href{https://zoom3.net/sonificacions/ona-gravitacional.html}{https://zoom3.net/sonificacions/ona-gravitacional.html}. La p\`{a}gina web \'{e}s interactiva i permet escollir la massa de chirp que es vol escoltar.

\subsection{Efecte de lent gravitat\`{o}ria}
\label{fonament-lent}

Mentre aquestes ones gravitacionals viatgen, despr\'{e}s de ser emeses, pot ser que trobin algun objecte astrof\'{i}sic massiu (per exemple un altre forat negre o una gal\`{a}xia) abans d’arribar a la Terra. Com que la gravetat d’aquest objecte corba l’espai-temps, segons la Teoria de la Relativitat General \cite{einstein-15}, les ones gravitacionals es distorsionaran i la seva traject\`{o}ria es deflectir\`{a}. Aquest efecte de distorsi\'{o} i deflexi\'{o} s’anomena efecte de lent gravitat\`{o}ria. Si alguna ona gravitacional pateix aquest efecte i la detectem, la seva forma ser\`{a} diferent que la de la Fig.~\ref{fig:h-t-unlensed}. L'efecte de lent gravitat\`{o}ria \'{e}s rar (afectaria aproximadament un de cada $1000$ senyals en les actuals observacions \cite{ng-18,li-18,oguri-18,mukherjee-21,wierda-21}) i encara no s’ha observat als detectors \cite{hannuksela-19, LVK-O3a-lensing, LVK-O3ab-Lensing}. Malgrat que no 
n'haguem detectat, podem modelitzar qu\`{e} es podria observar quan en detectem una.

La distorsi\'{o} ser\`{a} diferent depenent dels seg\"{u}ents par\`{a}metres: 

\begin{itemize}

\item la massa de la lent $M_L$, 

\item la freq\"{u}\`{e}ncia de l'ona gravitacional $f$, que dependr\`{a} de la massa de la font, dels forats negres ($m_1$ i $m_2$), com descriv\'{i}em a  les Eqs.~\eqref{eq:f-tau},\eqref{eq:freq-ms}.

\item el desalineament de la font respecte la lent i l’observador, representat pel par\`{a}metre $y$. Quan $y=0$, la font, la lent i l'observador estan perfectament alineats. A mesura que $y$ augmenta, es van desalineant i la distorsi\'{o} deguda a l'efecte de lent gravitat\`{o}ria va disminuint. Est\`{a} definit com 
\begin{equation}
y=\frac{d_L}{d_S}\frac{\eta}{R_{\rm E}}.
\end{equation}
\'{E}s un factor geom\`{e}tric que cont\'{e} una combinaci\'{o} de dist\`{a}ncies i escales caracter\'{i}stiques. Com es pot veure a la Fig.~\ref{fig:lens-geom}, $d_L$ \'{e}s la dist\`{a}ncia a la que est\`{a} la lent, $d_S$ la dist\`{a}ncia a la que est\`{a} la font, $\eta$ \'{e}s la dist\`{a}ncia transversal de desalineament i $R_{\rm E}$ \'{e}s una escala caracter\'{i}stica del sistema (anomenat radi d'Einstein)\footnote{$\displaystyle{R_{\rm E}=\sqrt{\frac{4 G M_L}{c^2} \frac{d_L d_{LS}}{d_S}}}$, conegut com a radi d'Einstein, \'{e}s l'escala caracter\'{i}stica de la lent gravitat\`{o}ria. $d_{LS}$ \'{e}s la dist\`{a}ncia entre la lent i la font (tal com es mostra a la Fig.~\ref{fig:lens-geom}).}. A efectes pr\`{a}ctics \'{e}s \'{u}til imaginar que $d_L$, $d_S$ i $R_{\rm E}$ estan fixos, i va variant $\eta$. $y$ \'{e}s directament proporcional a $\eta$.

\begin{figure}
    \centering
    \includegraphics[width=\columnwidth]{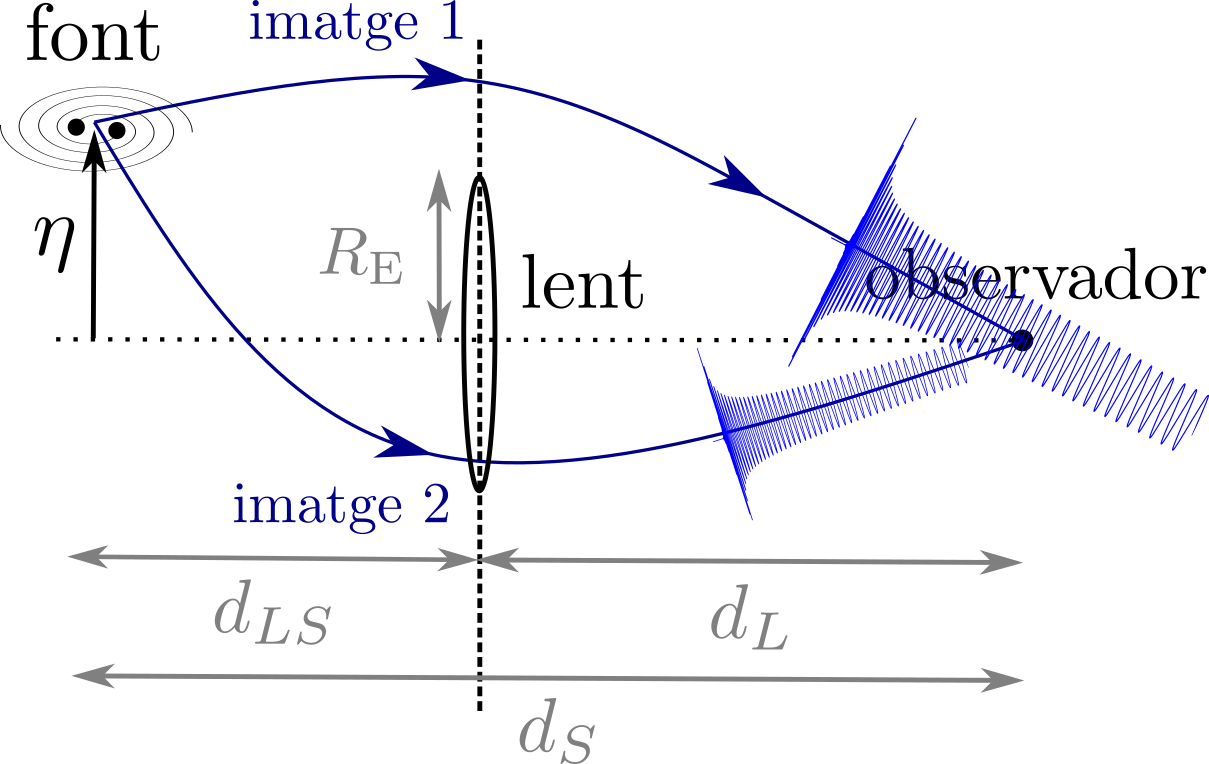}
    \caption{Esquema de l'efecte de lent gravitat\`{o}ria. Les ones gravitacionals de la font passen a trav\'{e}s d'una lent gravitat\`{o}ria (un altre objecte astrof\'{i}sic massiu), s\'{o}n distorsionades i deflectides, i arriben a l'observador (el detector a la Terra). En aquest cas, les l\'{i}nies blaves corbades representen la traject\`{o}ria de dues imatges que es podrien formar (en el model d'una massa ``puntual"), en l'aproximaci\'{o} d'\`{o}ptica geom\`{e}trica.}
    \label{fig:lens-geom}
\end{figure}

\'{E}s interessant que en molts casos el producte $f\cdot M_L$ actua com una variable independent\footnote{En alguns estudis es defineix una ``freq\"{u}\`{e}ncia adimensional" proporcional a $f\cdot M_L$.}. D'aquesta manera, per exemple, l'efecte que causaria disminuir $M_L$ seria el mateix que el que causaria disminuir $f$.

\end{itemize}

El senyal, quan es distorsiona, pot apar\`{e}ixer magnificat i/o multiplicat en diferents imatges, semblant al que ocorre en sistemes \`{o}ptics com lupes, lents i jocs de miralls. Hi ha principalment tres tipus d’efectes: (1) amplificaci\'{o} del senyal, (2) interfer\`{e}ncia entre m\'{u}ltiples imatges de la mateixa font, quan es superposen, (3) m\'{u}ltiples imatges separades de la mateixa font, quan no es superposen. 

Considerem que la lent gravitacional \'{e}s una massa ``puntual", \'{e}s a dir, concentrada en un punt de l'espai. Per a aquest model de lent, es formen dues imatges. Per a altres models de lents m\'{e}s complexes, pot ser que apareguin m\'{e}s de dues imatges.

El tipus d’efecte que puguem observar dep\`{e}n dels par\`{a}metres de la lent i de la geometria del sistema. Prendrem $y$ i el producte \mbox{$f\cdot M_L$} com a par\`{a}metres. Vegem qu\`{e} passaria si considerem un dels par\`{a}metres com a constant i anem variant l'altre:

\begin{itemize}

\item Primer, prenem el par\`{a}metre $y$ com a una constant. Si el producte $f\cdot M_L$ \'{e}s petit, hi haur\`{a} nom\'{e}s amplificaci\'{o} (Fig.~\ref{fig:h-t-lensed} (a)). Si aquest producte augmenta (per exemple, augmentant la massa de la lent, o augmentant la freq\"{u}\`{e}ncia de la font), llavors es pot tenir les dues imatges solapades, interferint (Fig.~\ref{fig:h-t-lensed} (b)). Si $f\cdot M_L$ augmenta encara m\'{e}s, aquestes imatges es veuran com a dos senyals separats (Fig.~\ref{fig:h-t-lensed} (c)).

\begin{figure}
    \centering
    \includegraphics[width=\columnwidth]{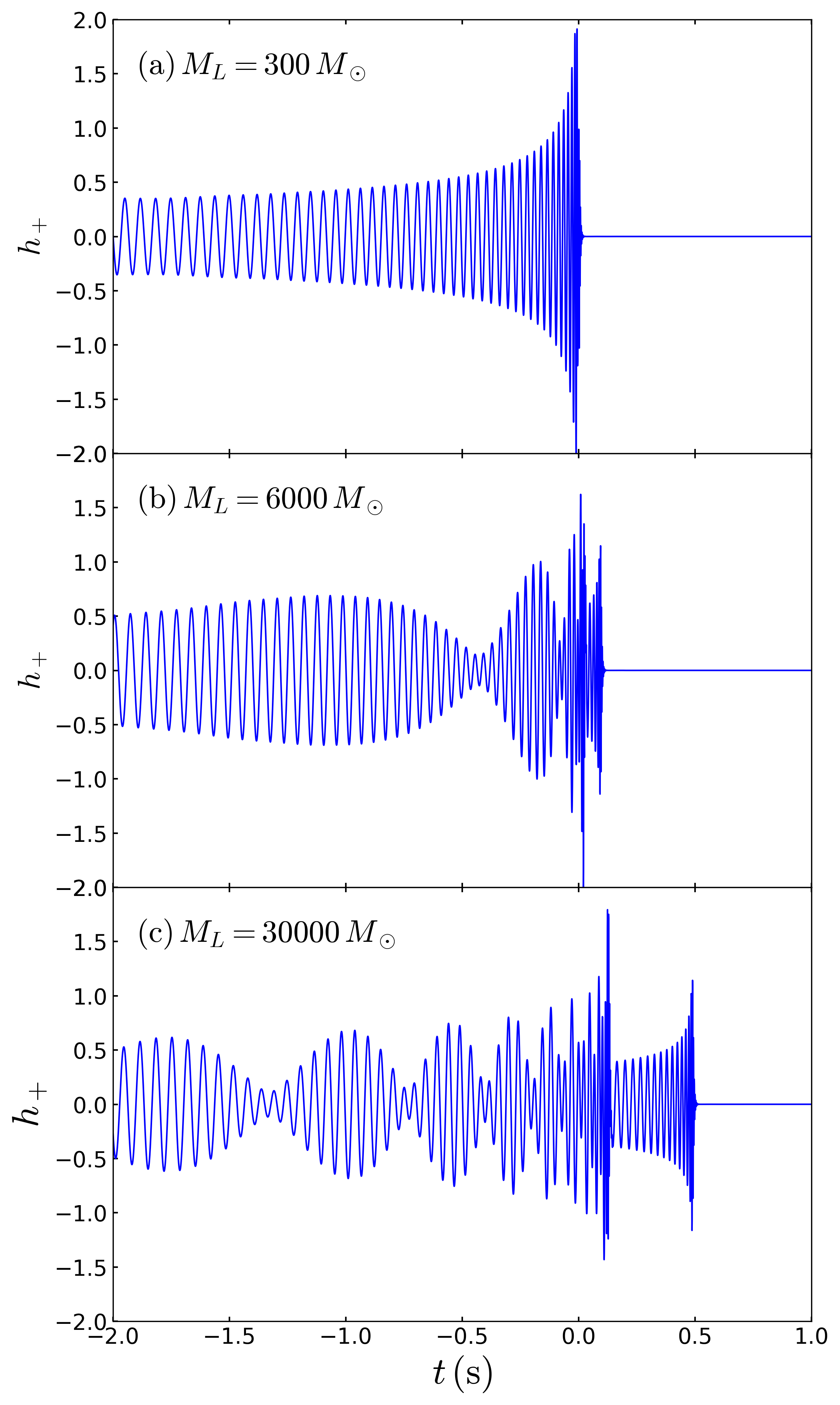}
    \caption{Deformaci\'{o} de l'espai-temps degut a una ona gravitacional afectada per l'efecte de lent gravitat\`{o}ria, $h_+^L$, en funci\'{o} del temps $t$. Els tres panells mostren diferents casos, amb el mateix valor de $y$ en tots tres ($y=0.3$). La massa de la lent va variant: (a) $M_L = 300 \, M_\odot$, (b) $M_L = 6000 \, M_\odot$, (c) $M_L = 30000 \, M_\odot$. Hem utilitzat alguns dels casos amb qu\`{e} es pot jugar a la p\`{a}gina web \href{https://zoom3.net/sonificacions/ona-gravitacional-lent.html}{https://zoom3.net/sonificacions/ona-gravitacional-lent.html}.}
    \label{fig:h-t-lensed}
\end{figure}

\item Per altra banda, ara prenem com a constants la massa de la lent i la freq\"{u}\`{e}ncia de la font (per tant $f\cdot M_L$ tamb\'{e} ser\`{a} constant). 
Si el par\`{a}metre $y$ \'{e}s un valor molt petit, les imatges encara no estan ben definides i podria haver-hi nom\'{e}s amplificaci\'{o}.
Si $y$ augmenta una mica, les imatges ja començaran a estar definides, per\`{o} estaran properes i solapades, i hi haur\`{a} interfer\`{e}ncia entre elles.
Per a $y$ majors, les imatges estan m\'{e}s separades. 

A m\'{e}s, l'amplificaci\'{o} de les imatges, com veurem ($\sqrt{\mu_1}$ i $\sqrt{\mu_2}$, Eqs.~\eqref{eq:mu1},\eqref{eq:mu2}) nom\'{e}s dep\`{e}n del par\`{a}metre $y$. Com m\'{e}s petit sigui $y$, m\'{e}s ``fort" es detectar\`{a} el senyal final (combinaci\'{o} de les dues imatges) en general.

\end{itemize}

A la p\`{a}gina web \href{https://zoom3.net/sonificacions/ona-gravitacional-lent.html}{https://zoom3.net/sonificacions/ona-gravitacional-lent.html} es pot jugar amb els par\`{a}metres $\mathcal{M}_c$, $M_L$ i $y$ per comprovar els diferents casos exposats.

\subsubsection{Aproximaci\'{o} d'\`{o}ptica geom\`{e}trica: m\'{u}ltiples imatges}

La distorsi\'{o} es pot quantificar amb l'anomenat ``factor de transmissi\'{o}" $\,F$ \cite{born-99,takahashi-03}. L'ona gravitacional amb l'efecte de lent es podr\`{a} descriure com a $h_+^{L} = F \cdot h_+$ (en l'espai de freq\"{u}\`{e}ncies).
Quan les imatges estan definides i apareixen separades o interferint, es pot utilitzar una aproximaci\'{o}, d'``\`{o}ptica geom\`{e}trica" (GO) \cite{schneider-92,nakamura-deguchi-99,takahashi-03}:
\begin{align}
h_+^{L} 
&= F_{GO} \cdot h_+ \nonumber \\
&= h_+ \sqrt{\mu_1}\,e^{i \phi_1} + h_+ \sqrt{\mu_2} \,e^{i \phi_2-i \pi/2}.
\label{eq:F_GO_PML}
\end{align}
Com podem observar, el senyal amb l'efecte de lent, $h_+^L$, \'{e}s la suma de dos senyals: el primer senyal \'{e}s com l'original $h_+$, per\`{o} l'amplitud es modifica amb el factor $\sqrt{\mu_1}$ (Eq.~\eqref{eq:mu1}); el segon senyal amb el factor $\sqrt{\mu_2}$ (Eq.~\eqref{eq:mu2}). A m\'{e}s, els dos v\'{e}nen amb una fase diferent (un amb $\phi_1$ i l'altre amb $\phi_2-\pi/2$). Si representem l'amplitud en termes de fase relativa entre les imatges, 
\begin{equation}
h_+^{L} 
= \left(h_+ \sqrt{\mu_1} + h_+ \sqrt{\mu_2} \,e^{i 2 \pi \Delta t - i \pi/2} \right) e^{i \phi_1},
\label{eq:GO-relative}
\end{equation}
on hem escrit $\phi_2-\phi_1 = 2\pi \Delta t$. Aix\`{o} significa que el segon senyal ve amb un desfasament de \mbox{$2 \pi \Delta t - \pi/2$} respecte el primer, que es tradueix en un retard temporal $\Delta t$ \footnote{Addicionalment, hi ha el terme $-\pi/2$ que prov\'{e} de la topologia de la funci\'{o} que determina el retard temporal.}. Per tant, el segon senyal arribar\`{a} un temps $\Delta t$ m\'{e}s tard que el primer. Aquesta separaci\'{o} temporal entre les imatges ve donada per
\begin{equation}
\Delta t \simeq 4\cdot 10^{-5} \,y \, \left(
\frac{M_L}{M_\odot}\right) \, {\rm s},
\label{eq:time-delay}
\end{equation}
en unitats de segons (s).
Com ens mostra l'expressi\'{o} matem\`{a}tica, el temps de separaci\'{o} (o retard) entre les imatges augmenta quan (i) augmenta $y$, (ii) augmenta $M_L$.

L’amplificaci\'{o}/magnificaci\'{o} de cada imatge nom\'{e}s dep\`{e}n del par\`{a}metre $y$. La magnificaci\'{o} de la primera i segona imatge s\'{o}n, respectivament, \cite{schneider-92}:
\begin{equation}
\sqrt{\mu_{1}}=\displaystyle{\sqrt{\frac{1}{4} 
\left( \frac{y}{\sqrt{y^2+4}} + \frac{\sqrt{y^2+4}}{y} + 2 \right) }},
\label{eq:mu1}
\end{equation}
\begin{equation}
\sqrt{\mu_{2}}=\displaystyle{\sqrt{\frac{1}{4} 
\left( \frac{y}{\sqrt{y^2+4}} + \frac{\sqrt{y^2+4}}{y} - 2 \right) }}.
\label{eq:mu2}
\end{equation}

L'aproximaci\'{o} d'\`{o}ptica geom\`{e}trica, per\`{o}, ja no \'{e}s v\`{a}lida en el cas on hi ha nom\'{e}s amplificaci\'{o}. All\'{i} hi ha difracci\'{o} i les imatges encara no estan totalment definides. La Fig.~\ref{fig:h-t-lensed} t\'{e} en compte l'efecte de difracci\'{o}. Per simplicitat, per\`{o}, per a la sonificaci\'{o} treballarem com si les imatges ja estiguessin definides sempre, utilitzant l'Eq.~\eqref{eq:F_GO_PML}. Per tant, cal tenir-ho en compte quan interpretem els resultats. L'aproximaci\'{o} d'\`{o}ptica geom\`{e}trica fa que l'amplificaci\'{o} es sobreestimi en el cas on nom\'{e}s hi ha amplificaci\'{o} (Fig.~\ref{fig:h-t-lensed} (a)), on la difracci\'{o} \'{e}s important. Per altra banda, no afecta als altres casos, d'interfer\`{e}ncia i imatges separades, on \'{e}s v\`{a}lida l'aproximaci\'{o} d'\`{o}ptica geom\`{e}trica. Si es vol tenir en compte correctament la difracci\'{o}, cal utilitzar un formalisme matem\`{a}tic m\'{e}s elaborat. Per a m\'{e}s detalls, es pot consultar la refer\`{e}ncia \cite{BU-2022}.

\subsubsection{Efectes d'ona: interfer\`{e}ncia i difracci\'{o}}

A m\'{e}s de produir m\'{u}ltiples imatges, la lent gravitat\`{o}ria tamb\'{e} pot produir efectes associats a la naturalesa d'ona de les ones gravitacionals. Aquests ``efectes d'ona" s\'{o}n principalment la interfer\`{e}ncia i la difracci\'{o}. 

La interfer\`{e}ncia (entre ones) es produeix quan dues o m\'{e}s ones es troben, i es genera una distorsi\'{o} en l'ona resultant. Per exemple, es pot veure si es tiren dues pedres a un estany: la interfer\`{e}ncia es produir\`{a} quan les ones es trobin. De la mateixa manera, hem vist que la lent gravitat\`{o}ria pot crear m\'{u}ltiples imatges de la mateixa font. Encara que les imatges estiguin separades en el temps $\Delta t$, com que tenen una certa durada, es podrien solapar. Quan aquestes imatges es solapen, es produeix interfer\`{e}ncia entre elles.  L'efecte de la interfer\`{e}ncia est\`{a} regit per la fase relativa de les ones: de l'Eq.~\eqref{eq:GO-relative}, veiem que la difer\`{e}ncia relativa de fase entre les dues imatges del nostre cas d'estudi seria $2\pi \Delta t - \pi/2$. En algunes regions la interfer\`{e}ncia ser\`{a} constructiva (les amplituds de les imatges es sumen) o destructiva (les amplituds de les imatges es resten). D'aquesta manera, es produir\`{a} un patr\'{o} d'interfer\`{e}ncia com a pulsacions (com es pot veure a la Fig.~\ref{fig:h-t-lensed} (b),(c)).

La difracci\'{o}, per altra banda, tamb\'{e} \'{e}s un fenomen d'interfer\`{e}ncia, per\`{o} de m\'{u}ltiples ones parcials (aqu\'{i} parlem d'ones en general, no d'imatges). Quan $\Delta t$ \'{e}s molt petit\footnote{Formalment, podem prendre la condici\'{o} de ``$\Delta t$ molt petit'' com $\Delta t \lesssim 1/f$.}, les dues imatges encara no estan ben definides. En aquest cas, la contribuci\'{o} prov\'{e} d'ones parcials (``parts de l'ona") de la font, distorsionades per la lent. La difracci\'{o} pot crear una amplificaci\'{o} del senyal (Fig.~\ref{fig:h-t-lensed} (a)). Si la comparem amb l'amplificaci\'{o} generada per l'efecte del solapament de les dues imatges, per\`{o}, l'amplificaci\'{o} deguda a la difracci\'{o} est\`{a} una mica m\'{e}s suprimida. Per a generar la Fig.~\ref{fig:h-t-lensed} (a) hem tingut en compte l'efecte de difracci\'{o}. No obstant, per a fer la sonificaci\'{o}, hem utilitzat el model de dues imatges definides (\`{o}ptica geom\`{e}trica). Aix\`{o} fa que per a valors petits del producte $f\cdot M_L\cdot y$ (quan nom\'{e}s hi ha amplificaci\'{o}), l'amplificaci\'{o} que veiem estar\`{a} sobrevalorada respecte al cas que la difracci\'{o} es tingu\'{e}s en compte. 

\subsection{Comparaci\'{o} de les ones gravitacionals amb el so}

Encara que sonifiquem les ones gravitacionals, no significa que es puguin escoltar en la realitat. El so s\'{o}n ones mec\`{a}niques (de pressi\'{o}) que nom\'{e}s poden viatjar a trav\'{e}s d’un medi 
material 
(aire/gas, per\`{o} tamb\'{e} a trav\'{e}s de l\'{i}quids i s\`{o}lids). La seva velocitat tamb\'{e} dep\`{e}n del medi a trav\'{e}s del qual es propaga. En canvi, les ones gravitacionals s\'{o}n ones del propi espai-temps del qual tots formem part, i es propaguen tant a trav\'{e}s del buit com de qualsevol medi a la velocitat de la llum. En aquest cas no \'{e}s l’ona que es propaga a trav\'{e}s de l’espai-temps, sin\'{o} la deformaci\'{o} del propi espai-temps que es propaga: quan ens travessa una ona gravitacional, nosaltres tamb\'{e} ens distorsionem perqu\`{e} estem dins de l'espai-temps (no cal patir, perqu\`{e} \'{e}s un efecte tant min\'{u}scul que no el notem). Malgrat les difer\`{e}ncies entre els dos tipus d'ones (so i ones gravitacionals), el propi comportament d’ona d’ambd\'{o}s en permet fer un s\'{i}mil prou prec\'{i}s.

\section{Proc\'{e}s de sonificaci\'{o}}

\subsection{Ones gravitacionals}

\subsubsection{Primera prova: correspond\`{e}ncia emp\'{i}rica}

El proc\'{e}s de sonificaci\'{o} s'ha realitzat en diverses fases. En una primera prova, s'ha realitzat la sonificaci\'{o} de manera aproximada. Per a fer-ho, ens hem basat en la forma de les gr\`{a}fiques creades a partir de les equacions. La intenci\'{o} d'aquesta part del proc\'{e}s \'{e}s obtenir un primer resultat sonor el m\'{e}s r\`{a}pidament possible.

En l'ona gravitacional que es mostra a la Fig.~\ref{fig:h-t-unlensed}, cal destacar la seva forma exponencial. La freq\"{u}\`{e}ncia tamb\'{e} augmenta de manera exponencial. Quan creem una ona exponencial gen\`{e}rica, al no provenir les dades de les equacions, podem escollir lliurement tots els valors fins a aconseguir una ona semblant: durada de l'ona, freq\"{u}\`{e}ncies inicials i finals, aix\'{i} com en el desfasament degut a l'efecte de lent gravitat\`{o}ria (seg\"{u}ent apartat, Sec.~\ref{sonif-lent}). Realitzant diverses iteracions, hem pogut replicar un patr\'{o} similar.

\subsubsection{Procediment realista}
\label{procediment-realista}

En un segon pas, hem utilitzat equacions realistes que descriuen les ones gravitacionals correctament. Hem generat dades de la variaci\'{o} exponencial de freq\"{u}\`{e}ncia usant l'Eq.~\eqref{eq:f-tau}.

Finalment hem utilitzat l'Eq.~\eqref{eq:h-t}, que genera l'ona gravitacional completa: freq\"{u}\`{e}ncia $f$ i amplitud $h_+$ en funci\'{o} del temps $t$, on fem servir $t = t_{\rm coal}-\tau$. Per evitar l'infinit a $\tau\rightarrow 0$, tallem l'equaci\'{o} a $\tau = 0.0001 \, {\rm s}$. D'aquesta manera, estem prenent la primera part de l'ona gravitacional com a senyal (des que els forats negres estan orbitant fins a aproximadament quan es tocarien i es fusionarien).  
L'Eq.~\eqref{eq:h-t}, a m\'{e}s de tenir en compte la variaci\'{o} de l'amplitud $h_+(t)$, tamb\'{e} inclou la variaci\'{o} de la freq\"{u}\`{e}ncia (Eq.~\eqref{eq:f-tau}) a trav\'{e}s de la fase $\Phi(\tau)$. 
La variaci\'{o} de freq\"{u}\`{e}ncia es pot observar a trav\'{e}s de la separaci\'{o} de les oscil·lacions a la Fig.~\ref{fig:h-t-unlensed}: comencen m\'{e}s separades (menor freq\"{u}\`{e}ncia) i van acostant-se (major freq\"{u}\`{e}ncia).

L'amplitud de tots els senyals s'ha escalat per a poder escoltar-los a un volum c\`{o}mode per a la nostra o\"{i}da. Cal anar amb compte en no tenir el volum molt alt quan els escoltem perqu\`{e}, encara que el senyal al principi se senti fluix, la intensitat del so augmentar\`{a} exponencialment. 

Les freq\"{u}\`{e}ncies de les ones gravitacionals que considerem s\'{o}n properes a les freq\"{u}\`{e}ncies de so audibles. 
No obstant, en alguns casos les freq\"{u}\`{e}ncies obtingudes queden fora del rang audible, que \'{e}s te\`{o}ricament de 20 Hz a 20000 Hz \footnote{Una altra limitaci\'{o} \'{e}s la resposta freq\"{u}\`{e}ncial dels altaveus de l'ordinador, que pot estar entre 80 Hz i 15000 Hz.}. Les ones gravitacionals provinents de forats negres m\'{e}s massius tindran freq\"{u}\`{e}ncies m\'{e}s baixes. Les freq\"{u}\`{e}ncies inicials seran tan greus que no les podrem sentir, nom\'{e}s sentirem la part final del senyal (on les freq\"{u}\`{e}ncies ja han augmentat prou per entrar al rang audible). Per a poder escoltar l'ona sencera, cal modificar les freq\"{u}\`{e}ncies obtingudes. Per tant, hem augmentat la freq\"{u}\`{e}ncia global de tots els senyals en $300$ Hz.

\subsection{Efecte de lent gravitat\`{o}ria}
\label{sonif-lent}

Per a simular l'efecte de lent gravitat\`{o}ria, hem utilitzat directament l'efecte realista, utilitzant l'aproximaci\'{o} d'\`{o}ptica geom\`{e}trica. 
L’efecte de lent gravitat\`{o}ria pot crear dues imatges de la mateixa font
\footnote{Com hem vist a la secci\'{o} \ref{fonament-lent}, hi ha una regi\'{o} on la difracci\'{o} \'{e}s important, on les imatges encara no estan totalment definides. Per al proc\'{e}s de sonificaci\'{o}, per\`{o}, utilitzem l'aproximaci\'{o} d'``\`{o}ptica geom\`{e}trica" (Eq.~\eqref{eq:F_GO_PML}) per simplicitat.}. Segons els par\`{a}metres de la lent, $M_L$ i $y$,
pot ser que les imatges arribin amb una difer\`{e}ncia de temps $\Delta t$ m\'{e}s curta o m\'{e}s llarga (Eq.~\eqref{eq:time-delay}). 
Per a la sonificaci\'{o},
enlloc de representar l'amplitud $h_+^L$ amb l'Eq.~\eqref{eq:F_GO_PML} directament (que es podria fer), hem decidit superposar dues c\`{o}pies del senyal original per a  simular l'efecte de lent. Cada senyal s'ha escalat amb el factor corresponent ($\sqrt{\mu_1}$ (Eq.~\eqref{eq:mu1}) pel primer, $\sqrt{\mu_2}$ (Eq.~\eqref{eq:mu2}) pel segon), i s'ha endarrerit el segon senyal en un factor $\Delta t$. Tamb\'{e} s'ha tingut en compte la fase extra $-\pi/2$ del segon senyal.

\subsubsection{Patr\'{o} d'interfer\`{e}ncia}

Per tal de crear les interfer\`{e}ncies, podem procedir de dues maneres:

1 - Reproducci\'{o} de les interfer\`{e}ncies de les ones de manera ac\'{u}stica, usant els altaveus esquerra i dret per a reproduir la mateixa ona desfasada, corresponent a dues imatges separades per un retard $\Delta t$. Les interfer\`{e}ncies es produeixen de manera natural com un fenomen ac\'{u}stic real, no calculades. El resultat dep\`{e}n, per tant, de la separaci\'{o} dels altaveus esquerra i dret, i tamb\'{e} de la posici\'{o} des d'on s'escolta.

2 - Si es vol obtenir un resultat m\'{e}s fidel, i que no depengui de la configuraci\'{o} del sistema de reproducci\'{o} i escolta, la creaci\'{o} de l'ona resultant es pot realitzar digitalment.

\'{E}s important remarcar que el patr\'{o} d'interfer\`{e}ncia es genera de manera natural, al superposar les dues ones desfasades en el temps.
Aix\`{o} \'{e}s gr\`{a}cies a que els senyals de so, al ser ones, tamb\'{e} tenen una fase.
La fase relativa entre els dos senyals reproduir\`{a} el
mateix patr\'{o} d'interfer\`{e}ncia que el que crearia l'efecte de lent gravitat\`{o}ria. 

\subsubsection{Desplaçament de les freq\"{u}\`{e}ncies}

Com coment\`{a}vem a l'apartat \ref{procediment-realista}, cal modificar les freq\"{u}\`{e}ncies obtingudes per a fer el senyal completament audible. Existeixen diverses estrat\`{e}gies per a fer-ho. 

La primera manera consisteix en modificar la velocitat de reproducci\'{o} del so obtingut. En aquest cas, cada cop que es multiplica la velocitat de reproducci\'{o} per dos, s'augmenta la freq\"{u}\`{e}ncia per dos, per\`{o} tamb\'{e} es redueix la durada del senyal a la meitat. 

Si volem mantenir la durada del senyal original i el patr\'{o} d'interfer\`{e}ncia, hi ha una altra manera que es basa en la transformada de Fourier (FFT, Fast Fourier Transform). Aix\`{o} ens permet desplaçar les components de la freq\"{u}\`{e}ncia obtingudes cap a la part d'altes freq\"{u}\`{e}ncies de l'espectre. Finalment, utilitzem la transformada inversa de Fourier (IFFT) per a recuperar l'ona. Aquest operaci\'{o} no modifica l'amplitud de l'ona i el patr\'{o} d'interfer\`{e}ncia es mant\'{e}. Aquesta operaci\'{o} pot distorsionar digitalment l'ona original, introduint artefactes no desitjats, i per tant el desplaçament de les freq\"{u}\`{e}ncies no pot ser massa gran.

\section{Resultats}

\subsection{Sonificaci\'{o} de les ones gravitacionals provinents de la fusi\'{o} de dos forats negres}

La sonificaci\'{o} de les ones gravitacionals provinents de la fusi\'{o} de dos forats negres es pot escoltar a la p\`{a}gina web:

\href{https://zoom3.net/sonificacions/ona-gravitacional.html}{https://zoom3.net/sonificacions/ona-gravitacional.html}

Podeu escollir la massa de chirp $\mathcal{M}_c$ de la font que desitgeu escoltar, escrivint-la en el requadre de text. Despr\'{e}s, premeu el bot\'{o} esquerre ``Send" per a crear la forma del senyal que hagueu triat, que es mostrar\`{a} en la imatge. Finalment, premeu el bot\'{o} dret ``Play sound" per a escoltar la sonificaci\'{o} corresponent.

Preguem que aneu amb compte amb el volum i el to dels sons. La massa de chirp per defecte $\mathcal{M}_c=30\, M_\odot$ \'{e}s una freq\"{u}\`{e}ncia mitjana segura, a partir de la qual un pot augmentar o disminuir-la gradualment.

\subsection{Sonificaci\'{o} de l'efecte de lent gravitat\`{o}ria en ones gravitacionals}

La sonificaci\'{o} de l'efecte de lent gravitat\`{o}ria en ones gravitacionals provinents de la fusi\'{o} de dos forats negres es pot escoltar a les p\`{a}gines web:

\href{https://zoom3.net/sonificacions/ona-gravitacional-lent.html}{https://zoom3.net/sonificacions/ona-gravitacional-lent.html}

\href{https://zoom3.net/sonificacions/ona-gravitacional-lent-exemples.html}{https://zoom3.net/sonificacions/ona-gravitacional-lent-exemples.html}

A la primera p\`{a}gina web, podeu escollir els valors de la massa de chirp $\mathcal{M}_c$ de la font, la massa de la lent $M_L$ i el par\`{a}metre $y$. La forma de l'ona gravitacional resultant es mostrar\`{a} avall. La sonificaci\'{o} interactiva, per\`{o}, encara est\`{a} en desenvolupament.

A la segona p\`{a}gina web, podeu escoltar algunes de les sonificacions gravades, mentre la primera p\`{a}gina web est\`{a} en desenvolupament.

\section{Discussi\'{o}}

La naturalesa d'ona del so ens permet reproduir l'efecte de lent gravitat\`{o}ria en ones gravitacionals. No nom\'{e}s les imatges individuals, sin\'{o} tamb\'{e} la interfer\`{e}ncia entre imatges, que es crea de manera natural. Podem comprovar que, encara que el so i les ones gravitacionals tinguin una naturalesa diferent, comparteixen comportaments i en podem fer \'{u}s per a la sonificaci\'{o}.

Per al proc\'{e}s de sonificaci\'{o}, ens hem pres la llibertat de desplaçar lleugerament el rang de freq\"{u}\`{e}ncies per tal que els senyals fossin audibles (cap a freq\"{u}\`{e}ncies m\'{e}s altes). Per a evitar modificar les altres caracter\'{i}stiques del senyal, sobretot el patr\'{o} d'interfer\`{e}ncia, hem utilitzat una m\`{e}tode amb la transformada de Fourier (FFT) i la seva inversa (IFFT). 

Hem utilitzat l'aproximaci\'{o} d'\`{o}ptica geom\`{e}trica, que considera que les imatges estan ben definides sempre. Aix\`{o} ens permet crear les dues imatges per separat i solapar-les, creant la interfer\`{e}ncia. Com hem comentat, aquesta aproximaci\'{o} perd validesa quan el desfasament temporal \'{e}s molt petit ($\Delta t \lesssim 1/f$), on apareix la difracci\'{o}. El principal efecte d'utilitzar l'aproximaci\'{o}, i no tenir en compte la difracci\'{o}, \'{e}s la sobreestimaci\'{o} de l'amplitud en el cas on l'efecte de lent nom\'{e}s provoca amplificaci\'{o} del senyal. Per la resta de casos (imatges m\'{u}ltiples i interfer\`{e}ncia), l'aproximaci\'{o} \'{e}s prou precisa.

\subsection{Futurs passos}

Les p\`{a}gines web estan en desenvolupament continu. Seguirem desenvolupant-les i actualitzant-les per a millorar la presentaci\'{o} i explicar-ne el funcionament.

Ara que hem fet la sonificaci\'{o} de l'efecte de lent gravitat\`{o}ria en ones gravitacionals, el seg\"{u}ent pas \'{e}s explorar noves sonificacions en la vessant m\'{e}s art\'{i}stica. El pas final d'aquest proc\'{e}s d'exploraci\'{o} i creaci\'{o} preveu ser una exposici\'{o} p\'{u}blica a Barcelona, amb un enfocament did\`{a}ctic.

\section*{Agra\"{i}ments} 

Agra\"{i}m als organitzadors de la iniciativa ``Sounds and Astrophysics" (``Arts $\&$ Science"), que ha promogut la col·laboraci\'{o} entre la Facultat de Belles Arts de la Universitat de Barcelona i l'Institut de Ci\`{e}ncies del Cosmos (ICCUB). L'Arnau Rios  i el Jordi Miralda van aportar recomanacions per a la idea inicial del projecte. Tamb\'{e} volem agra\"{i}r especialment l'Anna Argudo per la implicaci\'{o} i empenta en la organitzaci\'{o} del projecte. 

Per a generar les Figs.~\ref{fig:h-t-unlensed} i \ref{fig:h-t-lensed}, hem utilitzat el software PyCBC (\href{https://zenodo.org/records/10473621}{https://zenodo.org/records/10473621}). Per a crear les Figs.~\ref{fig:BH-merger} i \ref{fig:lens-geom}, hem utilitzat Inkscape. Per a realitzar les sonificacions, hem utilitzat SuperCollider.

L'Helena Ubach \'{e}s benefici\`{a}ria de l'ajut predoctoral FI-SDUR de la Generalitat de Catalunya. 

\end{otherlanguage}

\clearpage

\bibliography{sonification}

\end{document}